\documentclass[pdflatex,sn-mathphys-num]{sn-jnl}% Math and Physical Sciences Numbered Reference Style
\usepackage{graphicx}%
\usepackage{multirow}%
\usepackage{amsmath,amssymb,amsfonts}%
\usepackage{amsthm}%
\usepackage{mathrsfs}%
\usepackage[title]{appendix}%
\usepackage{xcolor}%
\usepackage{textcomp}%
\usepackage{manyfoot}%
\usepackage{booktabs}%
\usepackage{algorithm}%
\usepackage{algorithmicx}%
\usepackage{algpseudocode}%
\usepackage{listings}%
\theoremstyle{thmstyleone}%
\newtheorem{theorem}{Theorem}%  meant for continuous numbers
\newtheorem{proposition}[theorem]{Proposition}% 

\theoremstyle{thmstyletwo}%
\newtheorem{remark}{Remark}%

\theoremstyle{thmstylethree}%

\begin{document}

\title[Article Title]{Fisher-Rao distances between finite-energy signals in Gaussian noise}

%%=============================================================%%
%% GivenName	-> \fnm{Joergen W.}
%% Particle	-> \spfx{van der} -> surname prefix
%% FamilyName	-> \sur{Ploeg}
%% Suffix	-> \sfx{IV}
%% \author*[1,2]{\fnm{Joergen W.} \spfx{van der} \sur{Ploeg} 
%%  \sfx{IV}}\email{iauthor@gmail.com}
%%=============================================================%%

\author*[]{\fnm{Franck} \sur{Florin}} \email{}

%\author[2,3]{\fnm{Second} \sur{Author}}\email{iiauthor@gmail.com}
%\equalcont{These authors contributed equally to this work.}

%\author[1,2]{\fnm{Third} \sur{Author}}\email{iiiauthor@gmail.com}
%\equalcont{These authors contributed equally to this work.}

\affil[]{ \orgname{Thales}, \orgaddress{\country{France}}}

%\affil[2]{\orgdiv{Department}, \orgname{Organization}, \orgaddress{\street{Street}, \city{City}, \postcode{10587}, \state{State}, \country{Country}}}

%\affil[3]{\orgdiv{Department}, \orgname{Organization}, \orgaddress{\street{Street}, \city{City}, \postcode{610101}, \state{State}, \country{Country}}}

%%==================================%%
%% Sample for unstructured abstract %%
%%==================================%%

\abstract{This paper proposes representing finite-energy signals observed within a given bandwidth as parameters of a probability distribution and employing the information-geometric framework to compute the Fisher–Rao distance between these signals, considered as distributions. The observations are described by their discrete Fourier transforms, which are modelled as complex Gaussian vectors with known diagonal covariance matrices and parametrised means. These parameters define a coordinate system on a statistical manifold.
We investigate the possibility of deriving closed-form expressions for the Fisher–Rao distance. We employ established results from the Riemannian geometry of the multivariate normal model and extend the analysis to complex Gaussian variables representing the finite-energy signal observations. Expressions for the Christoffel symbols and the geodesic tensor equations in the Fisher metric are derived, leading to geodesic equations expressed as second-order differential equations. Although these equations depend on the parametric model, they  combine the magnitude and phase of the signal and their gradients with respect to the parameters. Two cases are examined: (1) the general case for any finite-energy signal observed in a given bandwidth and (2) the observation of a finite-energy signal with a known magnitude spectrum and unknown phases and attenuation coefficient. The manifold of finite-energy signals corresponds to the manifold of the multivariate normal model with a known diagonal covariance matrix, while the set of finite-energy signals with a known magnitude spectrum constitutes a submanifold. Closed-form expressions for the Fisher–Rao distances are obtained for both cases. We show that the submanifold is not geodesic, because the Fisher–Rao distance measured on the submanifold exceeds that on the entire manifold.}

\keywords{Fisher-Rao distance, Finite energy signal, Christoffel symbols, Fisher metric}

%%\pacs[JEL Classification]{D8, H51}

%%\pacs[MSC Classification]{35A01, 65L10, 65L12, 65L20, 65L70}

\maketitle

\section{Introduction}\label{sec1}

Time-series analysis and signal processing are well-established applications of information geometry. Modelling time series as realisations of statistical populations via parametric probability distributions enables the utilisation of tools from differential geometry.

Let $\boldsymbol{x}$ denote the observation vector and $\boldsymbol{\xi}$ the parameter vector. The stochastic behaviour of the observations is characterised by the conditional probability distribution $p(\boldsymbol{x} | \boldsymbol{\xi})$. As detailed in references \cite{bib_Amari_Nagaoka} and \cite{bib_amari_cichocki}, the set of all such parametric probability distributions forms a statistical model, which can be regarded as a statistical manifold. Given two probability distributions on the statistical manifold, we may define a divergence that measures their discrepancy. The manifold equipped with a divergence function has a differential-geometric structure. The geometry induced by the divergence function endows it with a Riemannian metric. Amongst the various Riemannian metrics, the Fisher metric is invariant with respect to the choice of coordinate system. It is also covariant under transformations of sufficient statistics.

The Fisher metric links the geometric structure of statistical manifolds with problems of statistical estimation. Within a Riemannian metric, a geodesic connecting two points is a curve along which the tangent vectors remain parallel (i.e. autoparallel transport). With the Fisher metric and the Levi-Civita connexion, the geodesic is furthermore the curve of minimal length connecting the two points. This minimal length defines a metric distance, known as the Fisher-Rao distance \cite{bib_nielsen2024}. Using the Fisher metric, one may evaluate this statistical distance between two points, characterised by their parameter vectors $\boldsymbol{\xi_1}$ and $\boldsymbol{\xi_2}$. This distance quantifies the dissimilarity between the two statistical populations described by the probability distributions $p(\boldsymbol{x} | \boldsymbol{\xi_1})$ and $p(\boldsymbol{x} | \boldsymbol{\xi_2})$ \cite{bib_Nielsen_ent}. The Fisher-Rao distance can be used for statistical analysis, including data clustering \cite{bib_strapasson} and deep learning, to name but a few \cite{bib_nielsen2023}.

When the parameters correspond to signal parameters, computing the Fisher-Rao distance between the associated distributions provides a measure of the dissimilarity between the corresponding signals, taking into account the distributions of the observations. A small distance implies that the distributions are similar and that the corresponding signals are difficult to distinguish on the basis of the observed data. In contrast, a large distance indicates that the signals are easily distinguishable.

With regard to signal analysis, stationary Gaussian time series are commonly examined via the statistical properties of their Fourier expansions. Amari and Nagaoka have derived the geometry of signal processing from the spectral density function, under the assumption that the sum of the squared logarithms of the power spectrum norm is finite \cite{bib_Amari_Nagaoka} (p. 116). Barbaresco investigated the information geometry of autoregressive (AR) models through their reflection coefficients \cite{bib_Barbaresco}. Choi and Mullhaupt established a correspondence between the information geometry of signal filters and Kähler manifolds \cite{bib_Choi}. In the conventional approach, the signal is modelled as a stationary time series subject to constraints on its power spectrum. These constraints describe the conditions under which the signal is produced as the output of a minimum-phase linear system, encompassing both AR and moving average (MA) models.
Centred circularly symmetric complex Gaussian vectors have been used in the context of radar with constant false alarm rate detection to model in-phase quadrature data and to apply the Siegel metric to evaluate a distance \cite{bib_lapuyade}. Arnaudon \textit{et al.} presented a geometric approach based on the Riemannian geometry of Toeplitz covariance matrices to design pulsed Doppler radar systems \cite{bib_arnaudon}. Distributed sensors have been considered by Sun \textit{et al}. to analyse complex cyber-physical systems. They developed a method to extract information from multivariate time series, based on covariance matrices and geodesic distances, to identify the nodes and edges of a network \cite{bib_sun}.
Radar target detection with a constant false alarm rate was also addressed by Ghojavand \textit{et al.} They compared various detectors based on Toeplitz covariance estimation for adaptive beamforming and detection \cite{bib_ghojavand}. To avoid degradation in detection performance caused by non-homogeneous clutter, Hua \textit{et al.} designed a detector within the framework of matrix information geometry, based on the total Bregman divergence on the Riemannian manifold of Hermitian Positive-Definite matrices \cite{bib_hua_2018} \cite{bib_hua_2021}.
Chen and Wu, working within the research group of the National University of Defence Technology, Changsha, China, used affine transformation and Bregman divergence to achieve satisfactory clutter-suppression performance in heterogeneous environments \cite{bib_chen} \cite{bib_wu_2022}. They used the Riemannian manifold of Hermitian positive-definite matrices as sample data correlation matrices. They further extended the method to a power-spectrum information geometry detector, which requires less computation \cite{bib_wu_2024}. In these approaches, the sample data represent radar cells, including target echoes with Doppler-frequency effects. The power spectrum is obtained after applying a digital Fourier transform to the cell data.

All these methods rely on the modelling of the raw signal as a stationary process and exploit either the covariance matrix or, following a Fourier expansion, the power spectrum. However, a wide range of applications, including telecommunications, sonar, radar, electronic warfare, music, speech analysis, fault diagnosis, and others, focus on the acquisition of time series representing finite-energy signals observed via noisy reception channels. In these contexts, the signal of interest often has not only finite energy but also finite duration, precluding its representation as a stationary random process. Consequently, the conventional approach is not suitable for modelling finite-energy signals. 

In this work, we propose an alternative approach to address the problem of parametric estimation for a finite-energy raw signal mixed with additive stationary noise. In the approach considered, the background noise is typically assumed to be stationary and Gaussian, whereas the signal is regarded deterministic and unknown. Each signal is associated with a parameter vector $\boldsymbol{\xi}$, which defines a unique probability distribution $p(\boldsymbol{x} | \boldsymbol{\xi})$ on the statistical manifold.

For conciseness, albeit at the risk of slight imprecision, the Fisher–Rao distance between the statistical distributions corresponding to two signals parameterised by $\boldsymbol{\xi}_1$ and $\boldsymbol{\xi}_2$ will hereafter simply be referred to as the distance between the two points, $\boldsymbol{\xi}_1$ and $\boldsymbol{\xi}_2$. Our objective is to compute the Fisher–Rao distance between two points, $\boldsymbol{\xi}_1$ and $\boldsymbol{\xi}_2$, on the statistical manifold. In the literature, some methods are devoted to approximate computation of the Fisher–Rao distance. For instance, in \cite{bib_collas}, the authors select intermediate points to compute divergences between non-centred signals using geodesic triangles. In \cite{bib_le_brigant}, geometric shooting is employed to approximate the minimum distance between curves, where the curves represent the evolution of stationarity for signals in the statistical manifold of centred stationary Gaussian processes. In this paper, we restrict our focus to the Fisher metric and to closed-form expressions of the Fisher–Rao distance, obtained as solutions to the geodesic equations.

Section \ref{sec2} presents the method of information geometry commonly used to compute the Fisher–Rao distance based on the Fisher metric. This section references previous work by the research community \cite{bib_Nielsen_ent} \cite{bib_Amari} \cite{bib_Pinele} \cite{bib_Miyamoto}. While it may not be necessary to re-explain the method in full detail, this section also provides an opportunity to introduce relevant definitions and notational conventions.  We recall the general approach for deriving the geodesic equations, as well as some results related to the multivariate normal distributions known to admit closed-form solutions. This section also explains why the geodesic equations (developed within the Fisher metric) have to be reconsidered to adapt conventional results to the finite-energy signal model.

Section \ref{sec3} presents the proposed finite-energy signal model and the associated statistical distributions. This signal model has previously been introduced in earlier work \cite{bib_Florin} \cite{bib_Florin_gretsi} \cite{bib_Florin_GSI25}. The parametric modelling is discussed in detail. We start with a generic approach that encompasses manifolds made of any parametrised finite-energy signals. Then we focus on two specific manifolds. One manifold includes all finite-energy signals in the observation bandwidth. The other manifold is a submanifold related to finite-energy signals with a known magnitude spectrum.

Section \ref{sec4} outlines the various steps required to compute the Fisher–Rao distance, including the formulation of the differential equations for geodesics. General formulas for the Fisher–Rao distance on the manifolds identified in section \ref{sec3} are provided. Full derivations with the calculation of Christoffel symbols are included in the appendices to facilitate understanding. This section also reviews key contributions from previous research on the Riemannian geometry of the multivariate normal model, with a particular focus on the computation of geodesic distances \cite{bib_nielsen2024} \cite{bib_nielsen2023} \cite{bib_Pinele} \cite{bib_skovgaard_81} \cite{bib_skovgaard_84}. Building upon these foundations, we introduce our novel approach, which extends these results to the context of finite-energy signal modelling in the presence of additive noise, a scenario not fully addressed in the existing literature.

Section \ref{sec5} evaluates the Fisher-Rao distances on the previously defined manifolds. We compare the distances on the entire manifold and on the submanifold and examine the asymptotic behaviours. We use numerical applications to compare the distances and analyse whether the submanifold of finite-energy signals with a known magnitude spectrum is fully geodesic on the entire manifold of all finite-energy signals. Figures are drawn to illustrate the numerical behaviour with some examples of signals. A numerical example is provided, demonstrating that the submanifold is not a fully geodesic submanifold; that is, the Fisher–Rao distance measured on the submanifold exceeds the corresponding distance on the entire manifold.

The conclusion summarises the main findings, discusses the relevance and limitations of the work, and outlines directions for future research.

\section{How to compute the Fisher-Rao distance}\label{sec2}

\subsection{Notational conventions} \label{subsec21}

Before describing the methodology for computing the Fisher–Rao distance, we introduce the notational conventions that will be employed throughout the paper.

Vectors and matrices are denoted in bold, while scalar values are presented in a regular typeface. For example, $\boldsymbol{[0]}_{P}$ represents the zero vector with $P$ components. The upper T denotes the transpose operation as in the expression $\boldsymbol{a}^{T}$, where $\boldsymbol{a}$ is a matrix or a vector.

$Re\left\lbrace c \right\rbrace$ and $Im\left\lbrace c \right\rbrace$ design, respectively, the real and imaginary parts of the complex number $c$. $c=Re\left\lbrace c \right\rbrace + \imath \cdot Im\left\lbrace c \right\rbrace$ with $\left( \imath\right) ^{2}=-1$. The upper * denotes the conjugate operation as in the expression $\boldsymbol{c}^{*}=Re\left\lbrace c \right\rbrace - \imath \cdot Im\left\lbrace c \right\rbrace$.

We adopt Einstein’s summation convention to omit summation signs in tensorial expressions, which allows a concise representation of tensor operations. However, when summing over indices, such as frequency, where the indexation is not strictly tensorial, we retain the explicit summation symbol, as in equation~\eqref{eq13}.

With regard to parameters, it is important to distinguish between $\boldsymbol{\xi}_2$, which refers to a particular value of the parameter vector $\boldsymbol{\xi}$, and $\xi^2$, which denotes the second component of $\boldsymbol{\xi}$.

To avoid any ambiguity between exponentiation and the upper index $2$, exponents are written in explicit parentheses; for example, $\big(\xi^2\big)^2$ denotes the square of the component $\xi^2$.

Where appropriate, the vector $\boldsymbol{\xi} = \begin{pmatrix} \boldsymbol{\phi} \\ \boldsymbol{\varphi} \end{pmatrix}$ can be decomposed into its subcomponents $\boldsymbol{\phi}$ and $\boldsymbol{\varphi}$, each with their respective indices:
\begin{itemize} \item $\boldsymbol{\xi}$ is a real vector of dimension $N$; \item $\boldsymbol{\phi}$ is a real vector of length $P$ indexed by $u$ or $v$, $1\leq u, v \leq P$; \item $\boldsymbol{\varphi}$ is a real vector with $(N-P)$ components indexed by $q$ or $r$, $P+1\leq q, r \leq N$. \end{itemize} This convention facilitates a Schouten-like notation for differentiation: $\partial_{u}=\frac{\partial}{\partial \phi^{u}}$, $\partial_{v}=\frac{\partial}{\partial \phi^{v}}$, $\partial_{q}=\frac{\partial}{\partial \varphi^{q}}$ and $\partial_{r}=\frac{\partial}{\partial \varphi^{r}}$.
\\
The gradient with respect to the parameters is, by definition, the operator: $ \boldsymbol{\bigtriangledown_{\boldsymbol{\xi}}}=\left[\frac{\partial}{\partial\xi^i}\right]_{ i \in \{1,2...N \}}$. By the definition of $\boldsymbol{\xi}$, $\boldsymbol{\phi}$ and $\boldsymbol{\varphi}$, we have:\\
\begin{align*}
   &\boldsymbol{\bigtriangledown_{\boldsymbol{\phi}}}=\left[\frac{\partial}{\partial\phi^u}\right]_{ u \in \{1,2...P \}} \\
   &\boldsymbol{\bigtriangledown_{\boldsymbol{\varphi}}}=\left[\frac{\partial}{\partial\varphi^q}\right]_{ q \in \{P+1,P+2...N \}} \\
   &\boldsymbol{\bigtriangledown_{\xi}}=\begin{pmatrix} \boldsymbol{\bigtriangledown_{\boldsymbol{\phi}}}\\
       \bigtriangledown_{\boldsymbol{\varphi}}
   \end{pmatrix}
\end{align*}

In a receiver, signals are often subject to time delay and attenuation. We denote the attenuation coefficient (assumed to be independent of both time and frequency) by $\alpha$, and the time delay by $\tau$, which is assumed to be constant over the frequency band since no Doppler effect is assumed.
\\

\subsection{Methodology to compute the Fisher-Rao distance}

The Fisher-Rao distance computation follows the methodology described in several foundational references~\cite{bib_Amari_Nagaoka, bib_Nielsen_ent, bib_Amari} and is widely used in the information geometry community~\cite{bib_Pinele, bib_Miyamoto}. The steps are as follows.

The first step is to specify a parametric model that describes the distribution of the observations ( $ p(\boldsymbol{x} \mid \boldsymbol{\xi})$ ).

The Fisher information matrix $\left[ g_{ij} \right]$ is derived from the log-likelihood as follows:
\begin{equation}
g_{ij} = E_{\boldsymbol{\xi}}\left[ \frac{\partial}{\partial \xi^i} \ln p(\boldsymbol{x} | \boldsymbol{\xi})
\frac{\partial}{\partial \xi^j} \ln p(\boldsymbol{x} | \boldsymbol{\xi}) \right] \label{eq01}
\end{equation}

Based on the Fisher information matrix, the Christoffel symbols are then computed according to:
\begin{equation}
\forall \ m,i,j = 1,...,N \ \ \ \      \Gamma_{ij,m} := g_{mk} \Gamma_{ij}^{k} = \frac{1}{2} \left( \frac{\partial g_{jm}}{\partial \xi^i} +
\frac{\partial g_{mi}}{\partial \xi^j} - \frac{\partial g_{ij}}{\partial \xi^m} \right) \label{eq02}
\end{equation} 
Once Christoffel symbols are known, they can be employed in the following differential equations, whose solutions correspond to the geodesics $\boldsymbol{\tilde{\xi}}(\varsigma)$ joining $\boldsymbol{\xi_1} = \boldsymbol{\tilde{\xi}}(0)$ and $\boldsymbol{\xi_2} = \boldsymbol{\tilde{\xi}}(1)$ (with $\varsigma \in \left[ 0,1 \right]$):
\begin{equation}
\forall \ m = 1,...,N \ \ \     g_{mk}\frac{d^2 \xi^k}{d\varsigma^2} + \Gamma_{ij,m} \frac{d \xi^i}{d\varsigma} 
\frac{d \xi^j}{d\varsigma} = 0 \label{eq03}
\end{equation}

Given the expressions of geodesics $\boldsymbol{\tilde{\xi}}(\varsigma)$, the Fisher-Rao distance between the distribution at $\boldsymbol{\xi_1} = \boldsymbol{\tilde{\xi}}(0)$ and that at $\boldsymbol{\xi_2} = \boldsymbol{\tilde{\xi}}(1)$ is given by the following integral:
\begin{equation}
d(\boldsymbol{\xi_1},\boldsymbol{\xi_2}) = \int_{0}^{1}
\sqrt{g_{ij}\frac{d \tilde{\xi}^i}{d\varsigma} \frac{d \tilde{\xi}^j}{d\varsigma}} d\varsigma \label{eq04}
\end{equation} \\

\subsection{The case of the multivariate normal distributions}

Finding closed-form expressions for the Fisher–Rao distance is recognised as a non-trivial endeavour, and such expressions are only available for a limited number of families of probability distributions \cite{bib_Miyamoto}. 
For example, if we consider the multivariate normal distributions $\mathcal{N}(\boldsymbol{\mu},\boldsymbol{\Sigma})$ given by :
\begin{equation}
p(\boldsymbol{X} | \boldsymbol{\mu}, \boldsymbol{\Sigma})=\dfrac{1}{2^{\frac{n}{2}} \pi^{\frac{n}{2}} \sqrt{ \det \left( \boldsymbol{\Sigma}\right) }} 
\exp \left(  -\dfrac{ \left( \boldsymbol{X}-\boldsymbol{\mu} \right)^T \boldsymbol{\Sigma}^{-1} \left( \boldsymbol{X}-\boldsymbol{\mu} \right)}  {2}  \right) 
\label{eq08}
\end{equation}

where:
\begin{itemize}
\item $n$ is an integer, the number of components,\\
\item $ \boldsymbol{X} $ is a real random vector with $n$ components,\\
\item $\boldsymbol{\Sigma}$ is a $n\times n$ positive-definite matrix,\\
\item $\boldsymbol{\mu}$ is a real vector with $n$ components.
\end{itemize}

The geodesic equations corresponding to the multivariate normal distributions $\mathcal{N}(\boldsymbol{\mu},\boldsymbol{\Sigma})$ take the form of second-order ordinary differential equations (with $\varsigma \in [0,1]$) \cite{bib_skovgaard_81} \cite{bib_skovgaard_84}\cite{bib_Pinele} :
\begin{equation}
\begin{cases}
  \frac{d^2\boldsymbol{\mu}}{d\varsigma^2}- \frac{d\boldsymbol{\Sigma}}{d\varsigma} \boldsymbol{\Sigma}^{-1} \frac{d\boldsymbol{\mu}}{d\varsigma}=\boldsymbol{[0]_{N}}\\
  \frac{d^2\boldsymbol{\Sigma}}{d\varsigma^2}+\frac{d\boldsymbol{\mu}}{d\varsigma}\frac{d\boldsymbol{\mu}^T}{d\varsigma}
  - \frac{d\boldsymbol{\Sigma}}{d\varsigma} \boldsymbol{\Sigma}^{-1} \frac{d\boldsymbol{\Sigma}}{d\varsigma}=\boldsymbol{[0]_{N \times N}}
\end{cases}
\label{eqn06}
\end{equation} 
These equations have been solved by fixing the initial value conditions in $\boldsymbol{\mu}$ and $\frac{d\boldsymbol{\mu}}{dt}$ \cite{bib_nielsen2024} \cite{bib_calvo}. However, these geodesic equations were not solved in the general case for boundary conditions, that is, fixing $\boldsymbol{\mu}$ at $\varsigma=0$ and $\varsigma=1$ ($\boldsymbol{\mu}(0)=\boldsymbol{\mu}_1$ and $\boldsymbol{\mu}(1)=\boldsymbol{\mu}_2$).

The solutions for the equations with the boundary conditions are known for the distribution $\mathcal{N}(\boldsymbol{\mu},\boldsymbol{\Sigma})$ in the following cases \cite{bib_Pinele}: 
\begin{itemize}
\item $\boldsymbol{\Sigma}$ is constant, or
\item $\boldsymbol{\mu}$ is constant, or
\item $\boldsymbol{\Sigma}$ is diagonal, or
\item $\boldsymbol{\Sigma}$ is diagonal and $\boldsymbol{\mu}$ is an eigenvector of $\boldsymbol{\Sigma}$.
\end{itemize}

As stated above, in the general case, there is no explicit closed-form expression for the Fisher–Rao distance between multivariate normal probability distributions. Nevertheless, a closed-form expression can be derived in particular instances, such as multivariate normal distributions that share a common covariance matrix. In this scenario, the Fisher–Rao distance reduces to the Mahalanobis distance between the means of the multivariate normal distributions \cite{bib_Pinele}. That is, when we consider the distribution $\mathcal{N}(\boldsymbol{\mu},\boldsymbol{\Sigma})$ with $\boldsymbol{\Sigma}$ constant and when we have $\boldsymbol{\xi}=\boldsymbol{\mu}$, we can write :
\begin{equation}
d(\boldsymbol{\xi_1},\boldsymbol{\xi_2})  = d(\boldsymbol{\mu_1},\boldsymbol{\mu_2})  =\sqrt{ \left( \boldsymbol{\mu}_{2}-\boldsymbol{\mu}_{1} \right)^T \boldsymbol{\Sigma}^{-1} \left( \boldsymbol{\mu}_{2}-\boldsymbol{\mu}_{1}\right)}  
\label{eqn07}
\end{equation}
For more general cases, the distance generally does not admit an explicit closed form and must be approximately evaluated with numerical techniques such as geodesic shooting or bound approximations \cite{bib_nielsen2024}.\\

\subsection{Fisher-Rao distance in a submanifold} \label{subsec24}

Constraints in the parametric model of probability distributions restrict the distributions to a submanifold, and the Fisher–Rao geodesic distance on the submanifold is generally larger than the Fisher–Rao geodesic distance measured on the global manifold. An example is given in \citep{bib_Pinele}, where it is shown that the Mahalanobis distance between the means of two multivariate normal distributions with the same covariance matrix is larger than the Fisher-Rao distance between the same distributions assuming no constraint on the covariance matrix. This is due to the path between the distributions: it can be shortened when one does not assume that the covariance matrix is constant \cite{bib_Pinele}.

So, given a manifold and a submanifold within this manifold, we must consider that the Fisher–Rao geodesic distance on the submanifold is generally greater than on the manifold. When examining the distance between two points $\boldsymbol{\xi_1}$ and $\boldsymbol{\xi_2}$, we must carefully consider the constraints imposed by the parameters and the geometric structure of the submanifold they induce. For example, equations \ref{eqn06} and \ref{eqn07} for the distribution $\mathcal{N}(\boldsymbol{\mu},\boldsymbol{\Sigma})$ with $\boldsymbol{\Sigma}$ constant follow the hypothesis that each component of the vector $\boldsymbol{\mu}$ is a coordinate of the submanifold \cite{bib_skovgaard_81}. This means that no relation links these coordinates and the components of the vector $\boldsymbol{\mu}$ vary freely along the geodesic curve. (By "freely, we mean that the number of degrees of freedom is equal to the number of coordinates $n$ of $\boldsymbol{\mu}$.) 

In the following, we provide a model for the acquisition of a finite-energy signal through a noisy receiving channel. The model determines the statistical distributions of the observations. As will be explained in detail, the observations are complex vectors, and the parameters remain real vectors. The signal model has an impact on the geometrical structure of the corresponding manifold. Based on this model, we redevelop the equations to obtain the Christoffel symbols and calculate a new version of the geodesic equations. In particular, we use the block diagonal structure of the Fisher matrix, obtained from the signal model, to calculate the simplified equations. We compare the new equations with the classical results. We show that, with additional hypotheses, in some cases, the geodesic equations can be solved to obtain a closed-form expression of the Fisher-Rao distance.

\section{Signal parametric modelling}\label{sec3}

\subsection{Observation of a finite-energy signal}\label{subsec31}

The model used to derive the expression of the statistical distribution has already been applied in previous works  \cite{bib_Florin,bib_Florin_gretsi,bib_Florin_GSI25}. We detail here the hypotheses and give the expression for the law of probability of the observations and its dependence on the parameters.

We are interested in applications recording finite-energy signals. As examples, we can name the following applications with some of the recorded signals:
\begin{itemize}
    \item Sonar or radar interception: Short-duration Radio Frequency (RF) pulses, chirp Frequency-Modulated (FM) signals, sonar pings,
    \item Passive acoustics : door slamming, glass breaking, bell clangs, thunderclaps, gunshots, bird calls, frog croaks, sperm whale clicks, machinery events (sudden spikes from arcing, valve closure impacts, beeps, buzzers),
    \item Telecommunications: bursts, data frames, symbol waveforms,
    \item Music: notes, chords.
\end{itemize}

In the considered applications, at the sensor level, the received signal $s_r(t)$ is a real signal with finite energy, depending on time $t$, for which we have $\int_{-\infty}^{+\infty} \left( s_r(t) \right)^2 dt < \infty$. This signal admits a Fourier expansion $\hat{s}_r(\nu)$, depending on frequency $\nu$. In the real world, the signal is only observed during a limited time interval: $t \in [0,T]$ and most of the time it has a finite duration.

We examine the observations after sampling and applying the Fourier transform.
Because, for real signals, the Fourier transform values at negative frequencies are the complex conjugates of the Fourier transform values at positive frequencies, we limit the observations to the positive frequencies.
The sensor has a limited bandwidth $\check{B}$, which is supposed to be a continuous interval in $\mathbb{R}^{+}$.
Due to the limited frequency bandwidth $\check{B}$ of the sensor, time sampling effect and Discrete Fourier Transform (DFT), the observation bandwidth is limited to a subset $\mathcal{B}$ of positive frequencies: $\mathcal{B} \subset \check{B} \subset \mathbb{R}^{+}$. In practice, this subset $\mathcal{B}$ is made up of $N_{\mathcal{B}}$ equispaced discrete values $\mathcal{B}=\{ \nu_1, \nu_2, ... \nu_{N_{\mathcal{B}}}\}$.
After time sampling and DFT, and from the receiver point of view, the signal is parametrised by a real vector $\boldsymbol{\xi}$ and for frequency dependence, we note $s_{\boldsymbol{\xi}}(\nu)$. It is important to note here that the parameter model, as it is defined, is considered after time sampling and DFT, and that it takes into account the limited receiver observation bandwidth. It is known in classical signal theory that, due to the Paley-Wiener theorem, the Fourier transform of a function with compact support cannot be with compact support, except if it is the null function. However, this does not contradict the possible finite duration of the signal $s_r(t)$ at the receiver level. We keep in mind that the model takes into account the pre-processing in the receiver including sampling and DFT.
As a consequence of Parseval's theorem, we can write $\sum_{\nu \in \mathcal{B}} \left| s_{\boldsymbol{\xi}}(\nu) \right|^2 \delta \nu < \infty$ (with $\delta \nu = {B}/N_{\mathcal{B}} $ ; $B$ is here the length of the interval $\check{B}$ ).

In addition to the above hypotheses, we assume that the observation is composed of the signal mixed with additive stationary random noise. Thus, the observation takes the form: \begin{equation} \forall \nu \in \mathcal{B} \quad x(\nu) = s_{\boldsymbol{\xi}}(\nu) + n(\nu) \label{eq06} \end{equation}

In this equation, $x(\nu)$, $s_{\boldsymbol{\xi}}(\nu)$, and $n(\nu)$ are complex numbers. The total vector of observation $\boldsymbol{x}$ is made up of all complex variables $x(\nu)$ in the observation bandwidth: $\boldsymbol{x}=\left( x(\nu) \right)_ {\nu \in \mathcal{B} }$. This vector has a total of $N_{\mathcal{B}}$ complex coordinates, each corresponding to one of the frequency values in the bandwidth $\mathcal{B}=\{ \nu_1, \nu_2, ... \nu_{N_{\mathcal{B}}}\}$.

The signal is assumed to be deterministic, which means that each value of the parameter vector $\boldsymbol{\xi}$ corresponds to a single frequency function, $\nu \mapsto s_{\boldsymbol{\xi}}(\nu)$.

Given a time sampling rate of the original time series, the number of frequencies $N_{\mathcal{B}}$ depends on the integration time, which is assumed to be greater than the duration of the signal or sufficiently long to capture all the information required from the signal. It is also assumed that the integration time is sufficiently long to assume the asymptotic independence, circularity, and Gaussian behaviour of the noise samples.

Looking at the noise contributions in the frequency domain $n(\nu)$, we assume that the noise components, after the Fourier transform, $n(\nu)$, are centred circularly symmetric complex  Gaussian random variables, independent across frequencies. The Gaussianity is true if the noise is naturally Gaussian, but it is also asymptotically true for other types of noise distributions. As examples, more general hypotheses are mentioned in the following papers \cite{bib_Brillinger} \cite{bib_Peligrad}. The asymptotic independence for any pair of frequencies is also reported in \cite{bib_Amari_Nagaoka}.

We assume that the noise spectral power density is known:
$\forall \nu \in \mathcal{B} \ \ \ \gamma_{0}(\nu)= E \left( n(\nu) n^{*}(\nu) \right)$ (where $^*$ designs the conjugated complex number). The hypothesis that the Gaussian complex noise is circularly symmetric means: $E \left( n(\nu) n(\nu) \right)=0$. Assuming that the noise spectral power density is known is not a strong limitation for many applications. Generally, the noise regime is assessed during periods when the signal is absent. This is a common signal processing practice for detection and classification. In the bandwidth, the values $\forall \nu \in \mathcal{B} \ \ \gamma_{0}(\nu)$ are known.

The statistical distributions are determined by the signal parameters and the noise characteristics. Assuming all previous hypotheses, the observations $\boldsymbol{x}$ depend on the parameters through a parametric law of probability, where the parameters characterise the signal. The law of the total vector of the observations $\boldsymbol{x}=\left( x(\nu) \right)_ {\nu \in \mathcal{B} }$ can be expressed as follows: 
\begin{equation}
p(\boldsymbol{x} | \boldsymbol{\xi})=\prod\limits_{\nu \in \mathcal{B}} \dfrac{1}{\pi \gamma_{0}(\nu)} 
\exp \left(  -\dfrac{ \left|  x(\nu)-s_{\boldsymbol{\xi}} (\nu)   \right|^2   }{\gamma_{0}(\nu)}  \right)
\label{eq07}
\end{equation}

The total vector of observation $\boldsymbol{x}$ has a total number $N_{\mathcal{B}}$ of complex components and can be rewritten as a real vector $\boldsymbol{X}$ with $2N_{\mathcal{B}}$ real components. With this notation, equation $\ref{eq07}$ is similar to the normal multivariate distribution expressed in \citep{bib_Pinele} and in equation \ref{eq08}:

\begin{equation}
p(\boldsymbol{X} | \boldsymbol{\xi})=\dfrac{1}{2^{N_{\mathcal{B}}} \pi^{N_{\mathcal{B}}} \sqrt{ \det \left( \boldsymbol{\Sigma}\right) }} 
\exp \left(  -\dfrac{ \left( \boldsymbol{X}-\boldsymbol{\mu}_{\boldsymbol{\xi}} \right)^T \boldsymbol{\Sigma}^{-1} \left( \boldsymbol{X}-\boldsymbol{\mu}_{\boldsymbol{\xi}} \right)}  {2}  \right) 
\label{eq08b}
\end{equation}

where:\\
\begin{itemize}

\item $N_{\mathcal{B}}$ is the number of frequencies $\nu$ in $\mathcal{B}$,\\
\item $\boldsymbol{\Sigma}=diag \left(\sigma_{0}^2 (n) \right)$ is a diagonal $2N_{\mathcal{B}}\times 2N_{\mathcal{B}}$ matrix, such that:
\begin{center}
$\forall k=1,..N_{\mathcal{B}}\ \ \   \sigma_{0}^2(2k-1)= \sigma_{0}^2(2k)=\frac{\gamma_{0} (\nu_k)}{2}$,\\
\end{center}
\item $ \boldsymbol{X} $ is a $2N_{\mathcal{B}}$ vector, with $ \forall k=1,...N_{\mathcal{B}} \ \ X^{2k-1}=Re\left\lbrace {x(k)}\right\rbrace  $ and  $ X^{2k}=Im \left\lbrace {x(k)} \right\rbrace  $,\\
\item $\boldsymbol{\mu}_{\boldsymbol{\xi}}$ is a $2N_{\mathcal{B}}$ vector, with $\forall k=1,...N_{\mathcal{B}} \ \ \mu_{\boldsymbol{\xi}}^{2k-1}=Re\left\lbrace s_{\boldsymbol{\xi}}\left( k \right) \right\rbrace  $
and  $\mu_{\boldsymbol{\xi}}^{2k}=Im \left\lbrace s_{\boldsymbol{\xi}}\left( k \right)  \right\rbrace $.\\
\end{itemize}

\subsection{Signal dependence on parameters and manifold definitions}\label{subsec32}

Equation $\ref{eq07}$ (or equivalently equation $\ref{eq08b}$) defines a family of probability distributions $S_{\Xi}$ when the parameter $\boldsymbol{\xi}$ varies in a predefined set ${\Xi}$:
\begin{equation}
S_{\Xi}=\{ p_{\boldsymbol{\xi}} = p(\boldsymbol{x} | \boldsymbol{\xi}) ~|~\boldsymbol{\xi} \in {\Xi} \}
\label{eq08a}
\end{equation}

Furthermore, we assume that the model verifies the following regularity conditions:
\begin{itemize}
\item $\Xi$ is an open subset of $\mathbb{R}^{2N_{\mathcal{B}}}$.
\item The mapping $\boldsymbol{\xi} \mapsto p_{\boldsymbol{\xi}}$ is injective.
\item And for any $\boldsymbol{x} \in \mathbb{C}^{N_{\mathcal{B}}}$ the function $\boldsymbol{\xi} \mapsto p(\boldsymbol{x} | \boldsymbol{\xi})$ is $\mathcal{C}^\infty$. (We could adopt hypothesis $\mathcal{C}^2$ or even hypothesis $\mathcal{C}^1$, but in practical cases most models verify $\mathcal{C}^\infty$.)
\end{itemize}
With these hypotheses, we may consider $S_{\Xi}$ as a statistical manifold (see \cite{bib_Amari_Nagaoka}, section 2.1 for more details).
As a consequence, the geometry of the manifold lies in equation $\ref{eq07}$ (or in equation $\ref{eq08b}$) and in the definition of the parameter set ${\Xi}$.\\

The general case concerns a situation where $\boldsymbol{\xi}$ has $N$ components, with $N \leq 2N_{\mathcal{B}}$, and the previous regularity conditions are verified. We call such a manifold, manifold $L2(\mathcal{B},\boldsymbol{\xi})$. As $\boldsymbol{\xi}$ has $N$ components, $N$ is the dimension of the manifold $L2(\mathcal{B},\boldsymbol{\xi})$.

For all $\nu$ in $ \mathcal{B} $, the signal $s_{\boldsymbol{\xi}} (\nu)$ is a complex value, which can be represented with a modulus (or magnitude) $\rho_{\boldsymbol{\xi}} (\nu)$ and a phase $\psi_{\boldsymbol{\xi}}(\nu)$ : 
\begin{equation}
s_{\boldsymbol{\xi}} (\nu)= \rho_{\boldsymbol{\xi}} (\nu) \cdot \exp \left(\imath \psi_{\boldsymbol{\xi}} (\nu) \right)
\label{eq09}
\end{equation}
With $\psi_{\boldsymbol{\xi}}(\nu)  \in \left] -\pi, \pi\right]$, and $\rho_{\boldsymbol{\xi}} (\nu) \in \left]0 , +\infty\right[$.\\

With expression \ref{eq09}, we see that the dependence of the signal on the parameters affects magnitude and phase simultaneously. Each signal is characterised by its magnitude and phase as functions of frequency. As functions of $\nu$, $\rho_{\boldsymbol{\xi}} (\nu)$ and $\psi_{\boldsymbol{\xi}} (\nu) $ are respectively called magnitude spectrum and phase spectrum. They are respectively the magnitude and the phase values of the signal Fourier transform.

Except when there is an additional constraint, such as the minimum phase, in general, a signal cannot be uniquely determined solely by the phase or the magnitude of its Fourier transform \citep{bib_Quatieri}. Indeed, we assume that, for the recovery of a signal, phase and magnitude spectra can be combined freely. That is, any phase function ${\psi}_{\boldsymbol{\xi}}(\nu)$ can be combined with any magnitude function $\rho_{\boldsymbol{\xi'}} (\nu)$ with $\boldsymbol{\xi}$ and $\boldsymbol{\xi'}$ being values arbitrarily chosen in $\Xi$. This allows us to particularise the model in the following way without loss of generality.

The parameters $\boldsymbol{\xi}$ of the signal are split into two parts: $\boldsymbol{\xi} = \begin{pmatrix} \boldsymbol{\phi} \\ \boldsymbol{\varphi} \end{pmatrix} $. One part, $\boldsymbol{\phi}$, is related to the magnitude $\rho_{\boldsymbol{\phi}} (\nu)$ of the signal, and the other part, $\boldsymbol{\varphi}$, is related to the phase $\psi_{\boldsymbol{\varphi}} (\nu)$ of the signal.

\begin{equation}
\forall \nu \in \mathcal{B} \ \ \ \ \ s_{\boldsymbol{\xi}} (\nu)= \rho_{\boldsymbol{\phi}} (\nu) \cdot \exp \left(\imath \psi_{\boldsymbol{\varphi}} (\nu) \right)     \label{eq10}
\end{equation} \\

In the following, any manifold $L2(\mathcal{B},\boldsymbol{\xi})$ is described by \ref{eq07} completed with equation \ref{eq10} which refers to the case where $\boldsymbol{\xi} = \begin{pmatrix} \boldsymbol{\phi} \\ \boldsymbol{\varphi} \end{pmatrix}$.\\
As the magnitudes and phases can be determined separately from the $s_{\boldsymbol{\xi}} (\nu)$, we assume that the numbers of coordinates of parameter vectors $\boldsymbol{\xi}$ verify the following properties:
\begin{equation}
\begin{cases}
    P \leq N_{\mathcal{B}} \\
    N-P \leq N_{\mathcal{B}} 
\end{cases}
    \label{eq10f}
\end{equation} \\
In a manifold $L2(\mathcal{B},\boldsymbol{\xi})$, the expression of the distribution is given by:
\begin{equation}
p(\boldsymbol{x}| \boldsymbol{\xi})=\prod\limits_{\nu \in \mathcal{B}} \dfrac{1}{\pi \gamma_{0}(\nu)} 
\exp \left(  -\dfrac{ \left|  x(\nu)-\rho_{\boldsymbol{\phi}} (\nu) \cdot \exp \left(\imath \psi_{\boldsymbol{\varphi}} (\nu)\right)     \right|^2   }{\gamma_{0}(\nu)}  \right)
\label{eq11}
\end{equation} 

We have to assume that the model satisfies the regularity conditions. $\Xi$ can be chosen to be an open subset of $\mathbb{R}^{2N_{\mathcal{B}}}$ and the mapping $\boldsymbol{\xi} \mapsto p_{\boldsymbol{\xi}}$ can be injective. In addition, the model must satisfy the hypothesis that for any $\boldsymbol{x} \in \mathbb{C}^{N_{\mathcal{B}}}$ the function $\boldsymbol{\xi} \mapsto p(\boldsymbol{x} | \boldsymbol{\xi})$ is $\mathcal{C}^\infty$. We can assume that the magnitude $\rho_{\boldsymbol{\phi}}(\nu)$ is $\mathcal{C}^\infty$ with respect to the parameters $\phi^u$. We will also assume that the phase $\psi_{\boldsymbol{\varphi}}(\nu)$ is $\mathcal{C}^\infty$ with respect to the parameters $\varphi^r$.

As we examine the signal in the observation bandwidth $ \mathcal{B} $, it is important to consider the dependence of the phase $\psi_{\boldsymbol{\varphi}}(\nu)$ on the frequency $\nu $. Although the observation bandwidth $\mathcal{B}$ is assessed only for a finite number $N_{\mathcal{B}}$ of values of $\nu$, it is possible to adopt a model valid for the continuous interval $\check{B} \supset \mathcal{B}$.

A discontinuity in the phase with respect to $\nu$ can occur when the phase reaches an extreme value of the interval, $-\pi $ or $ \pi $, causing the phase to jump to the other end of the interval. To reconstruct a continuous phase with respect to $\nu$, we can perform phase unwrapping. We denote the unwrapped phase by $\breve{\psi}_{\boldsymbol{\varphi}} (\nu)$. Consequently, the signal can be expressed as $ s_{\boldsymbol{\xi}} (\nu)= \rho_{\boldsymbol{\phi}} (\nu) \cdot \exp \left(\imath \breve{\psi}_{\boldsymbol{\varphi}} (\nu) \right)$.

So, the notation $\breve{\psi}_{\boldsymbol{\varphi}} (\nu)$ represents the unwrapped phase, which is continuous with respect to $\nu$. ${\psi}_{\boldsymbol{\varphi}} (\nu)$ represents the same phase in the interval $\left] -\pi, \pi\right]$ (i.e. modulo $2 \pi$). We denote this property as follows: ${\psi}_{\boldsymbol{\varphi}} (\nu)=\left\langle \breve{\psi}_{\boldsymbol{\varphi}} (\nu) \right\rangle_{\left] -\pi,\pi\right] }$.

The use of $\breve{\psi}_{\boldsymbol{\varphi}} (\nu)$ is especially interesting when we want to adopt continuous and smooth polynomial models with respect to $\nu$.

\subsection{The finite-energy signals and the manifold $L2(\mathcal{B})$}

The first case of specific manifold corresponds to the situation for which in equation $\ref{eq08b}$ the parameters are the components of the vector $\boldsymbol{\mu}_{\boldsymbol{\xi}}$. This means $\boldsymbol{\mu}_{\boldsymbol{\xi}}=\boldsymbol{\xi}$.

In this case $\boldsymbol{\xi}$ is a vector with $2N_{\mathcal{B}}$ components and $\forall k=1,...2N_{\mathcal{B}} \ \ \ \xi^k=\mu^{k}$. In other words: $\Xi = \mathbb{R}^{2N_{\mathcal{B}}}$.
This means that all the signal components $s_{\boldsymbol{\xi}}(\nu)$ vary freely in $\mathbb{C}$.

Variations in all signal components determine a manifold, which we call the manifold $L2(\mathcal{B})$. This corresponds to all signals with finite energy observed in the frequency bandwidth $\mathcal{B}$.
The dimension of the manifold $L2(\mathcal{B})$ is $2N_{\mathcal{B}}$. The manifold $L2(\mathcal{B})$ corresponds exactly to the manifold described by $2N_{\mathcal{B}}$ multivariate normal distributions with the same diagonal covariance matrix $\boldsymbol{\Sigma}$ and different mean values.\\

Any manifold $L2(\mathcal{B},\boldsymbol{\xi})$ is a submanifold of the manifold $L2(\mathcal{B})$.

\subsection{The finite-energy signals with a known magnitude spectrum and the manifold $L2(\mathcal{B},\alpha)$}

A common situation in the detection of finite-energy signals corresponds to the following case. The signal power spectrum (that is, the signal magnitude spectrum $\rho_{\boldsymbol{\xi}}(\nu) ~ \forall \nu \in \mathcal{B}$) is known except for a global attenuation coefficient $\alpha$, and the phase spectrum $\psi_{\boldsymbol{\xi}}(\nu) ~ \forall \nu \in \mathcal{B}$ remains unknown.

To address this situation, we particularise the parametric description and define the manifold $L2(\mathcal{B},\alpha)$, which is a specific case of manifold $L2(\mathcal{B},\boldsymbol{\xi})$. We have to make sure that this model verifies the regularity conditions.

We assume that the observation remains within the frequency interval $\check{B}=\left]\nu_0-B/2,\nu_0+B/2 \right[$. We suppose that the magnitude dependence $\rho_{\boldsymbol{0}} (\nu)~~ \forall \nu \in \check{B}$ is known, except for a global attenuation $\alpha$:
\begin{equation}
\forall \nu \in \check{B} ~~~~\rho_{\boldsymbol{\phi}} (\nu)=\alpha \cdot \rho_{\boldsymbol{0}} (\nu)
\label{eq19}
\end{equation}
Thus, there is only one parameter related to the magnitude: $P=1$ and $\phi^1=\alpha$.\\

In addition, we specify the model associated with the phase. The continuous phase is supposed to be polynomial for $\nu \in \check{B}$, that is:
\begin{equation}
\forall \nu \in \check{B} ~~~~\breve{\psi}_{\boldsymbol{\varphi}} (\nu)=\sum_{r=2}^{N} \varphi^r \cdot {(\nu)}^{r-2}
\label{eq20}
\end{equation}
We assume that the order of the polynomial, $N-2$, is lower than or equal to $N_{\mathcal{B}}-1$. It is sufficient to approximate all phase values $\psi(\nu_k)$ in $\mathcal{B}$ as precisely as needed. $\mathcal{B}$ contains exactly $N_{\mathcal{B}}$ frequency values. We  use the Lagrange method to create a polynomial of order lower than or equal to $N_{\mathcal{B}}-1$ that fits the $N_{\mathcal{B}}$ couples $(\nu_k,\psi(\nu_k))$.

\begin{remark}
The number $N-1$ of coefficients $\varphi^r$ of the polynomial function describing the phase allows the phase to be approximated as precisely as possible at all phase values in $\mathcal{B}$. This means that any observed phase function can be described by the model. Thus, the model applies to all signals with a known magnitude spectrum, an unknown attenuation, and an unknown phase spectrum.
\end{remark}

The phase of the model is the value of $\breve{\psi}_{\boldsymbol{\varphi}} (\nu)$ reduced modulo $2\pi$ in the interval $\left] -\pi,\pi\right]$.
We write: ${\psi}_{\boldsymbol{\varphi}} (\nu)=\left\langle \breve{\psi}_{\boldsymbol{\varphi}} (\nu) \right\rangle_{\left] -\pi,\pi\right] }$.

As the phase is reduced modulo $2\pi$, we limit the domain of definition of the coefficient $\varphi^2$ to the interval $\left] -\pi,\pi\right]$.

The second coefficient is related to the time delay $\tau$: $\varphi^3=-2 \pi \tau$.

The boundary conditions completing the geodesic equations are given by the known values of the parameters at the points $\boldsymbol{\xi_{1}}= \begin{pmatrix} {\alpha_1} \\ \boldsymbol{\varphi_1} \end{pmatrix}$ and $\boldsymbol{\xi_{2}}= \begin{pmatrix} {\alpha_2} \\ \boldsymbol{\varphi_2} \end{pmatrix}$, that are:\\ 
$\alpha \left( 0\right) =\alpha_{1}$,\\
$\psi_{\boldsymbol{\varphi}} \left( 0 \right) =\psi_{\boldsymbol{\varphi_{1}}}$,\\
$\alpha \left( 1 \right) =\alpha_{2}$,\\
$\psi_{\boldsymbol{\varphi}} \left( 1 \right) =\psi_{\boldsymbol{\varphi_{2}}}$.\\
$\psi_{\boldsymbol{\varphi_{2}}}$ and $\psi_{\boldsymbol{\varphi_{2}}}$ are polynomial functions of frequency $\nu$.

\begin{proposition}
   The  model of finite-energy signals with a known magnitude spectrum verifies the regularity conditions, proving that $L2(\mathcal{B},\alpha)$ is a manifold. 
\end{proposition}

\begin{proof} -\\
The number of parameters is $N$, with $P=1$.
The parameters belong to an open set:
\begin{itemize}
\item $\phi^1=\alpha \in ]0,+\infty[$, open set.\\
\item $\forall r\in \{3,...N\} ~~ \varphi^r \in \mathbb{R}$, open set.\\
\item $\varphi^2 \in \left] -\pi,\pi\right]$. It is not an open set. However, we can limit the variation in the open set $\left] -\pi,\pi\right[$ and handle the case $\varphi^2=\pi$ as a limit. Or we can map the domain locally with an interval $\varphi^2 \in \left] -\pi-\epsilon,\pi+\epsilon\right[$.
\end{itemize}
The mapping is injective:
\begin{itemize}
\item The mapping ${\phi^1} \mapsto \rho_{\boldsymbol{\phi}}(\nu)$ for $\nu \in \check{B}$ is injective.\\
\item The mapping $\boldsymbol{\varphi} \mapsto \breve{\psi}_{\boldsymbol{\varphi}} (\nu) $ for $\nu \in \check{B}$ is injective .
\end{itemize}
The distribution is $\mathcal{C}^\infty$with respect to the parameters:
\begin{itemize}
\item $\rho_{\boldsymbol{\phi}} (\nu)$ is $\mathcal{C}^\infty$ with respect to $\phi^1$ ; $\frac{\partial \rho_{\boldsymbol{\phi}}}{\partial \phi^1} (\nu)=\rho_0 (\nu) $ ; $\frac{\partial^2 \rho_{\boldsymbol{\phi}}}{\partial^2 \phi^1} (\nu)=0 $\\
\item $\breve{\psi}_{\boldsymbol{\varphi}} (\nu)$ is $\mathcal{C}^\infty$ with respect to any $\varphi^r ~~\forall r\in \{2,...N\}$ ; $\frac{\partial \breve{\psi}_{\boldsymbol{\varphi}}}{\partial \varphi^r} (\nu)= (\nu)^{r-2} $ ; $\frac{\partial^2 \breve{\psi}_{\boldsymbol{\varphi}}}{\partial^2 \varphi^r} (\nu)=0 $
\end{itemize}
To fully satisfy the regularity conditions, we need to extend the properties of $\breve{\psi}_{\boldsymbol{\varphi}} (\nu)$ to ${\psi}_{\boldsymbol{\varphi}} (\nu)$, especially:
\begin{itemize}
\item ${\psi}_{\boldsymbol{\varphi}} (\nu)$ is $\mathcal{C}^\infty$ with respect to any $\varphi^r ~~\forall r\in \{2,...N\}$
\item The mapping $\boldsymbol{\varphi} \mapsto {\psi}_{\boldsymbol{\varphi}} (\nu) $ for $\nu \in B$ is injective 
\end{itemize}
For $\nu$ in an open subset in $\check{B}$, we can write: $\exists k \in \mathbb{Z} ~~| ~~\breve{\psi}_{\boldsymbol{\varphi}} (\nu)={\psi}_{\boldsymbol{\varphi}}(\nu) +2 k \pi$.
This shows that the partial derivability properties with respect to $\boldsymbol{\varphi}$ are the same for $\breve{\psi}_{\boldsymbol{\varphi}} (\nu)$ and ${\psi}_{\boldsymbol{\varphi}} (\nu)$ (except for $\nu$ in a set of measure zero).

The $N-1$ Lagrange polynomials constitute a basis of the vector space of polynomials of degree at most $N-2$. Per construction, the unwrapped phase with a given coefficient $\varphi^2$ can also be uniquely determined. There is a bijection between ${\psi}_{\boldsymbol{\varphi}} (\nu)$ and $\breve{\psi}_{\boldsymbol{\varphi}} (\nu) $. So the mapping $\boldsymbol{\varphi} \mapsto {\psi}_{\boldsymbol{\varphi}} (\nu) $ for $\nu \in B$ remains injective.

\end{proof}

\section{General expressions of the Fisher-Rao distances} \label{sec4}

The objective of this section is to use the Fisher metric to calculate the geodesic equations and solve these equations to calculate closed-form expressions of the Fisher-Rao distances on the manifold $L2(\mathcal{B})$, on any submanifold $L2(\mathcal{B},\boldsymbol{\xi})$ and on the submanifold $L2(\mathcal{B},\alpha)$. 

\subsection{Geodesic equations on the manifold $L2(\mathcal{B})$}

The geodesic equation on the manifold $L2(\mathcal{B})$ is derived directly from equation \ref{eqn06} and equation \ref{eq08b}, when the coefficients of the matrix $\boldsymbol{\Sigma}$ are not part of the coordinates. In this case the second line of equation \ref{eqn06} does not exist, and the geodesic equation becomes:

\begin{equation}
\begin{cases}
  \frac{d^2\boldsymbol{\mu}_{\boldsymbol{\xi}}}{d\varsigma^2}=\boldsymbol{[0]_N}\\
 \end{cases}
\label{eqn08}
\end{equation}

The solution between $\boldsymbol{\mu_{\xi_1}}$ and $\boldsymbol{\mu_{\xi_2}}$ is an affine function of $\varsigma$ \citep{bib_Pinele}:
\begin{equation}
    \forall \varsigma \in [0,1]  ~~~~ \boldsymbol{\tilde{\mu}}_{\boldsymbol{\xi}}(\varsigma)=\boldsymbol{\mu_{\xi_1}}+\varsigma \left( \boldsymbol{\mu_{\xi_2}}-\boldsymbol{\mu_{\xi_1}}\right)
    \label{eq16n}
\end{equation}

\subsection{Fisher-Rao distance on the manifold $L2(\mathcal{B})$ }

From equation \ref{eq08b} and the definition of the Fisher matrix \ref{eq01}, we can easily derive the expression of the Fisher matrix in the manifold $L2(\mathcal{B})$:
\begin{equation}
    g_{ij} = \frac{\partial \boldsymbol{\mu_{\boldsymbol{\xi}}^T} }{\partial \xi^i} \boldsymbol{\Sigma^{-1}} \frac{\partial \boldsymbol{\mu}_{\boldsymbol{\xi}} }{\partial \xi^j} 
\end{equation}
If we report this expression of $\boldsymbol{G}$ in equation \ref{eq04}, we have the expression of the Fisher-Rao distance on the manifold $L2(\mathcal{B})$:
\begin{equation}
d_{L2,\mathcal{B}}(\boldsymbol{\xi_1},\boldsymbol{\xi_2}) = \int_{0}^{1}
\sqrt{\frac{d \boldsymbol{\tilde{\mu}_{\boldsymbol{\xi}}^T}}{d\varsigma} \boldsymbol{\Sigma^{-1}} \frac{d \boldsymbol{\tilde{\mu}}_{\boldsymbol{\xi}}}{d\varsigma}} d\varsigma 
\label{eq18n}
\end{equation}

It is known from \citep{bib_Pinele} that in the $2N_{\mathcal{B}}$ dimensional manifold composed of multivariate normal distributions with a common covariance matrix $\boldsymbol{\Sigma}$, the Fisher-Rao distance between two distributions parametrised respectively by $\boldsymbol{\mu}_{\boldsymbol{\xi_{1}}}$ and $\boldsymbol{\mu}_{\boldsymbol{\xi_{2}}}$ is equal to the Mahalanobis distance. This can be easily verified here. Changing $\boldsymbol{\tilde{\mu}}$ in \ref{eq18n} by its value from \ref{eq16n}, we have:
\begin{equation}
d_{L2,\mathcal{B}}(\boldsymbol{\xi_1},\boldsymbol{\xi_2}) =\sqrt{ \left( \boldsymbol{\mu}_{\boldsymbol{\xi_{2}}}-\boldsymbol{\mu}_{\boldsymbol{\xi_{1}}} \right)^T \boldsymbol{\Sigma}^{-1} \left( \boldsymbol{\mu}_{\boldsymbol{\xi_{2}}}-\boldsymbol{\mu}_{\boldsymbol{\xi_{1}}} \right)}  
\label{eq12}
\end{equation}

With the notation of equations \ref{eq08b} and \ref{eq09}, this Fisher-Rao distance on the manifold $L2(\mathcal{B})$ can be rewritten as follows.
\begin{align}
& d_{L2,\mathcal{B}} \left(\boldsymbol{\xi_{1}},\boldsymbol{\xi_{2}}  \right) \notag \\
& =\sqrt{ \sum_{\nu \in \mathcal{B}} \dfrac {2} {\gamma_{0}\left( \nu \right)}  
	\left( \left( \rho_{\boldsymbol{\xi_{2}}} \left( \nu \right) \right)^2 
	+ \left( \rho_{\boldsymbol{\xi_{1}}} \left( \nu \right) \right)^2 
	- 2 \rho_{\boldsymbol{\xi_{2}}} \left( \nu \right) \cdot
	\rho_{\boldsymbol{\xi_{1}}} \left( \nu \right) \cdot \cos \left( \psi_{\boldsymbol{\xi_{2}}}\left( \nu \right) - \psi_{\boldsymbol{\xi_{1}}}\left( \nu \right) \right) \right) 	}  
\label{eq13}
\end{align}

\begin{proof}[Proof]
- \\
    The $2N_{\mathcal{B}}$ frequencies in $\mathcal{B}$ are $\nu_1$, $\nu_2$,..., $\nu_{N_{\mathcal{B}}}$. \\
    $\boldsymbol{\Sigma}$ is a diagonal $2N_{\mathcal{B}} \times 2N_{\mathcal{B}}$ real matrix. Each value $\nu_i$ in the frequency band is represented by two diagonal values: $\left(\sigma_0({2i-1})\right)^2$ and $\left(\sigma_0({2i})\right)^2$ and we have by definition:
   \begin{equation*}
       \left(\sigma_0({2i-1})\right)^2=\left(\sigma_0({2i})\right)^2=\frac{\gamma(\nu_{i})}{2}
   \end{equation*}
   \\
   The vector $\boldsymbol{\mu_\xi}$ has $2N_{\mathcal{B}}$ components grouped by two:
   \begin{equation*}
        \begin{cases}
       \mu_\xi^{2i-1}=Re\left\lbrace s_{\boldsymbol{\xi}}\left( \nu_i \right) \right\rbrace=\rho_{\boldsymbol{\xi}} (\nu_i) \cdot \cos \left(\psi_{\boldsymbol{\xi}} 
       (\nu_i) \right)
    \\
       \mu_\xi^{2i}=Im\left\lbrace s_{\boldsymbol{\xi}}\left( \nu_i \right) \right\rbrace=\rho_{\boldsymbol{\xi}} (\nu_i) \cdot \sin \left(\psi_{\boldsymbol{\xi}} 
       (\nu_i) \right)
       \end{cases}
   \end{equation*}
At frequency $\nu_{i}$, the contribution of the components indexed by $2i-1$ and $2i$ is as follows:
\begin{align*}
    &\frac {\left(\mu_2^{2i-1}-\mu_1^{2i-1}\right)^2}{\left(\sigma_0({2i-1})\right)^2}+\frac {\left(\mu_2^{2i}-\mu_1^{2i}\right)^2}{\left(\sigma_0({2i})\right)^2}\\
    &=\frac{2}{\gamma(\nu_{i})} \cdot \left( \left( \rho_{\boldsymbol{\xi}_2} (\nu_i) \cos \left(\psi_{\boldsymbol{\xi}_2}  (\nu_i) \right)-\rho_{\boldsymbol{\xi}_1} (\nu_i) \cos \left(\psi_{\boldsymbol{\xi}_1}  (\nu_i) \right)\right)^2 \right)\\
    &+\frac{2}{\gamma(\nu_{i})} \cdot \left( \left( \rho_{\boldsymbol{\xi}_2} (\nu_i) \sin \left(\psi_{\boldsymbol{\xi}_2}  (\nu_i) \right)-\rho_{\boldsymbol{\xi}_1} (\nu_i) \sin \left(\psi_{\boldsymbol{\xi}_1}  (\nu_i) \right)\right)^2 \right)\\
    &=\frac{2}{\gamma(\nu_{i})} \cdot \left( \left( \rho_{\boldsymbol{\xi}_2} (\nu_i) \right)^2 +\left( \rho_{\boldsymbol{\xi}_1} (\nu_i) \right)^2 \right)\\
    &- \frac{2}{\gamma(\nu_{i})} \cdot 2 \rho_{2} (\nu_i) \rho_{1}(\nu_i) \cdot \left(\cos \left(\psi_{2}  (\nu_i) \right) \cos \left(\psi_{1}  (\nu_i) \right)+\sin \left(\psi_{\boldsymbol{\xi}_2}  (\nu_i) \right) \sin \left(\psi_{\boldsymbol{\xi}_1}  (\nu_i) \right)\right) \\
    &=\frac{2}{\gamma(\nu_{i})} \cdot \left( \left( \rho_{\boldsymbol{\xi}_2} (\nu_i) \right)^2 +\left( \rho_{\boldsymbol{\xi}_1} (\nu_i) \right)^2 \right)- \frac{2}{\gamma(\nu_{i})} \cdot 2 \rho_{\boldsymbol{\xi}_2} (\nu_i) \rho_{\boldsymbol{\xi}_1}(\nu_i) \cos \left( \psi_{\boldsymbol{\xi}_2}  (\nu_i) -\psi_{\boldsymbol{\xi}_1}  (\nu_i)\right) \\
    &=\frac{2}{\gamma(\nu_{i})} \cdot \left( \left( \rho_{\boldsymbol{\xi}_2} (\nu_i) \right)^2 +\left( \rho_{\boldsymbol{\xi}_1} (\nu_i) \right)^2 -  2 \rho_{\boldsymbol{\xi}_2} (\nu_i) \rho_{\boldsymbol{\xi}_1}(\nu_i) \cos \left( \psi_{\boldsymbol{\xi}_2}  (\nu_i) -\psi_{\boldsymbol{\xi}_1}  (\nu_i)\right) \right)
\end{align*}
   
\end{proof}

We take the example of two points with a known magnitude spectrum and unknown attenuations and phases : $\boldsymbol{\xi_{1}}= \begin{pmatrix} {\alpha_1} \\ \boldsymbol{\varphi_1} \end{pmatrix}$ and $\boldsymbol{\xi_{2}}= \begin{pmatrix} {\alpha_2} \\ \boldsymbol{\varphi_2} \end{pmatrix}$.

The Fisher-Rao distance between $\boldsymbol{\xi_1}$ and $\boldsymbol{\xi_2}$ on the manifold $L2(\mathcal{B})$ takes the form:
\begin{equation}
d_{L2,\mathcal{B}} \left(\boldsymbol{\xi_{1}},\boldsymbol{\xi_{2}}  \right) 
=\sqrt{\omega_{0}} \cdot \sqrt{  \left( \alpha_2\right)^2 
	+ \left( \alpha_1 \right)^2- 2  \alpha_2 \cdot	\alpha_1 \cdot {\frac{1}{\omega_{0}} \sum_{\nu \in \mathcal{B}} \dfrac{2}{\gamma_{0}(\nu)} \left(\rho_{0}(\nu)\right)^2 \cos\left(\Delta \psi (\nu) \right) }}
\label{eq24}\\
\end{equation}
With:\\
$\omega_{0}=\sum_{\nu \in \mathcal{B}} \dfrac{2}{\gamma_{0}(\nu)} \left(\rho_{0}(\nu)\right)^2$\\
$\Delta \psi (\nu)=\psi_{\boldsymbol{\varphi_{2}}}(\nu)-\psi_{\boldsymbol{\varphi_{1}}}(\nu)$

\begin{proof} - \\
Use equation \ref{eq19} in equation \ref{eq13}.\\
\end{proof}

\subsection{Geodesic equations on the submanifold $L2(\mathcal{B},\xi)$}

So far, we have analysed geodesics and Fisher-Rao distance in the manifold $L2(\mathcal{B})$ that encompasses all finite-energy signals in the observed bandwidth $\mathcal{B}$. We have not placed any constraints on the signal, which can freely evolve on the manifold. We now introduce additional constraints on the spectrum of the signal and investigate the implications of incorporating these constraints into the signal model. The constraints are represented by a parametric vector $\boldsymbol{\xi}$ with a limited number of components. The number of components ($N$) is less than or equal to that of the vector $\boldsymbol{X}$ ($2N_{\mathcal{B}}$). As mentioned in section \ref{subsec32}, these constraints define the submanifold $L2(\mathcal{B},\xi)$.

To compute the geodesic equations on the submanifold $L2(\mathcal{B},\xi)$, we use the method described in Section \ref{sec2} and the equations \ref{eq01}, \ref{eq02}, \ref{eq03}, and \ref{eq11}. The calculations are detailed in the appendices.

\begin{theorem} [Linearly Dependant Gradients (LDG) theorem] \label{th1}
The (Fisher-)geodesic equations in the submanifold $L2(\mathcal{B},\xi)$ reduce to the following two vectorial equations:
\begin{equation}
\begin{cases}
\sum_{\nu \in \mathcal{B}} \dfrac{2 \left(\rho_{\boldsymbol{\xi}}(\nu)\right)^2}{\gamma_{0} (\nu)}  
\left( \frac{d^{2} \ln{\rho_{\boldsymbol{\xi}}} }{d\varsigma^{2}}(\nu) + \left(\frac{d \ln{\rho_{\boldsymbol{\xi}}} }{d\varsigma}(\nu)  \right)^2- \left(\frac{d \psi_{\boldsymbol{\xi}}}{d\varsigma}(\nu) \right)^2  \right) 
\boldsymbol{\bigtriangledown}_{\boldsymbol{\xi}} \ln{\rho_{\boldsymbol{\xi}}}(\nu) = \boldsymbol{\left[ 0\right]_{N}} 
\label{eq14n} \\

\sum_{\nu \in \mathcal{B}} \dfrac{2\left(\rho_{\boldsymbol{\xi}}(\nu)\right)^2}{\gamma_{0}(\nu)}  
\left( \frac{d^{2} \psi_{\boldsymbol{\xi}}}{d\varsigma^{2}}(\nu) 
+2  \frac{d \ln{\rho_{\boldsymbol{\xi}}}}{d\varsigma} (\nu) \frac{d \psi_{\boldsymbol{\xi}}}{d\varsigma}(\nu)   \right) 
\boldsymbol{\bigtriangledown}_{\boldsymbol{\xi}} \psi_{\boldsymbol{\xi}}(\nu) = \boldsymbol{\left[ 0\right]_{N}}

\end{cases}
\end{equation}
\end{theorem}

\begin{proof} - \\
 
The appendix \ref{secA} gives the Fisher information matrix $\boldsymbol{G}$ and the partial derivatives $\partial_{i}\boldsymbol{G}$.
In the appendix \ref{secB}, we calculate the Christoffel symbols $\Gamma_{ij,k}$.
The geodesic equation is analysed and transformed in the appendix \ref{secC0}, starting from equation \ref{eq03} to obtain the following equations \ref{eq44s}.
\begin{equation}
\begin{cases}
\sum_{\nu \in \mathcal{B}} \dfrac{2}{\gamma_{0}(\nu)}  
\left( \frac{d^{2} \rho_{\boldsymbol{\phi}}}{d\varsigma^{2}}(\nu) -\rho_{\boldsymbol{\phi}}(\nu) \left(\frac{d \psi_{\boldsymbol{\varphi}}}{d\varsigma}(\nu) \right)^2  \right) 
\boldsymbol{\bigtriangledown}_{\boldsymbol{\phi}} \rho_{\boldsymbol{\phi}}(\nu) = \boldsymbol{\left[ 0\right]_{P}} 
\label{eq44s}  \\

\sum_{\nu \in \mathcal{B}} \dfrac{2}{\gamma_{0}(\nu)}  
\left(  \frac{d^{2} \psi_{\boldsymbol{\varphi}}}{d\varsigma^{2}}(\nu) 
+2 \rho_{\boldsymbol{\phi}}(\nu) \frac{d \rho_{\boldsymbol{\phi}}}{d\varsigma}(\nu)  \frac{d \psi_{\boldsymbol{\varphi}}}{d\varsigma} (\nu)  \right) 
\boldsymbol{\bigtriangledown}_{\boldsymbol{\varphi}} \psi_{\boldsymbol{\varphi}}(\nu) = \boldsymbol{\left[ 0\right]_{N-P}}
 \\
\end{cases}
\end{equation} \\

By definition, $\tilde{\rho}_{\boldsymbol{\xi}}=\tilde{\rho}_{\boldsymbol{\phi}}$ and $\tilde{\psi}_{\boldsymbol{\xi}}=\tilde{\psi}_{\boldsymbol{\varphi}}$.

Due to the derivative properties, we have $\frac{d {\tilde{\rho}_{\boldsymbol{\xi}}}}{d\varsigma}(\nu) ={\tilde{\rho}_{\boldsymbol{\xi}}(\nu)}\frac{d \ln{\tilde{\rho}_{\boldsymbol{\xi}}}}{d\varsigma}(\nu)$ and $\boldsymbol{\bigtriangledown}_{\boldsymbol{\xi}} {\tilde{\rho}_{\boldsymbol{\xi}}}(\nu) ={\tilde{\rho}_{\boldsymbol{\xi}}(\nu)}\boldsymbol{\bigtriangledown}_{\boldsymbol{\xi}}  \ln{\tilde{\rho}_{\boldsymbol{\xi}}}(\nu)$.

Because $\boldsymbol{\xi} = \begin{pmatrix} \boldsymbol{\phi} \\ \boldsymbol{\varphi} \end{pmatrix}$, we also have the following equalities:
\begin{align*}
    &\boldsymbol{\bigtriangledown}_{\boldsymbol{\xi}}=
    \begin{pmatrix} \boldsymbol{\bigtriangledown}_{\boldsymbol{\phi}} \\ 
        \boldsymbol{\bigtriangledown}_{\boldsymbol{\varphi}} 
        \end{pmatrix} \\
    &\boldsymbol{\bigtriangledown}_{\boldsymbol{\phi}}\psi_{\boldsymbol{\varphi}}(\nu)=\boldsymbol{\left[ 0\right]_{P}} \\
    &\boldsymbol{\bigtriangledown}_{\boldsymbol{\varphi}}\rho_{\boldsymbol{\phi}}(\nu)=\boldsymbol{\left[ 0\right]_{N-P}} \\
\end{align*}

Using these equalities, equation \ref{eq44s} is easily transformed into equation \ref{eq14n}.

\end{proof}
    
\begin{remark}
    When $\rho_{\boldsymbol{\phi}}(\nu)$ tends to zero, equations \ref{eq14n} remain valid because the quantities  $\left(\rho_{\boldsymbol{\xi}}(\nu)\right)^2 \frac{d^{2} \ln{\rho_{\boldsymbol{\xi}}} }{d\varsigma^{2}}(\nu)$ and $\left(\rho_{\boldsymbol{\xi}}(\nu)\right) \frac{d \ln{\rho_{\boldsymbol{\xi}}} }{d\varsigma}(\nu)$ tend to finite values.
\end{remark}

\begin{remark} [LDG theorem significance]
	These equations imply that, respectively, for the magnitude and the phase, the gradients with respect to the parameters, expressed for all different frequencies, are linearly dependent. There exists a linear combination of the gradients equal to zero, with coefficients that are not all zero. In itself, this property may not always mean much, because the numbers of coordinates of the gradients ($N$) are lower than the number of gradient vectors ($N_{\mathcal{B}}$). However, what is important in these expressions is that, for the geodesic, the coefficients are expressed as derivatives with respect to the curvilinear coordinate.
\end{remark}

\subsection{Fisher-Rao distance on the submanifold $L2(\mathcal{B},\xi)$} \label{subsec44}

\begin{proposition}
With $\tilde{\rho}_{\boldsymbol{\xi}} $ and $\tilde{\psi}_{\boldsymbol{\xi}}$ solutions of the (Fisher-)geodesic equations, the Fisher-Rao distance on the submanifold $L2(\mathcal{B},\xi)$ can be calculated as the following integral:
\begin{equation}
d_{L2,\mathcal{B},\boldsymbol{\xi}}(\boldsymbol{\xi_{1}},\boldsymbol{\xi_{2}}) = \int_{0}^{1}
\sqrt{ \sum_{\nu \in \mathcal{B}} \dfrac{2 \left(\tilde{\rho}_{\boldsymbol{\xi}}(\nu) \right)^2 }{\gamma_{0}(\nu)}  
	\left( \left( \frac{d \ln{\tilde{\rho}_{\boldsymbol{\xi}}}}{d\varsigma}(\nu) \right)^2
	+  \left(  \frac{d \tilde{\psi}_{\boldsymbol{\xi}}}{d\varsigma}(\nu) \right)^2\right)  }
d\varsigma \label{eq17}
\end{equation}
\end{proposition}

\begin{proof} -\\

We note $\tilde{\phi}^{u}$ (or $\tilde{\phi}^{v}$) and $\tilde{\varphi}^{q}$ (or $\tilde{\varphi}^{r}$) the solutions of the geodesic equations \ref{eq44s}.   They are expressed as functions of $\varsigma$.
From equation \ref{eq04} taking into account the block diagonal structure of the Fisher information matrix as demonstrated in appendix \ref{secA}, we derive the Fisher-Rao distance as the following integral:
\begin{equation}
d_{L2,\mathcal{B},\boldsymbol{\xi}}(\boldsymbol{\xi_{1}},\boldsymbol{\xi_{2}}) = \int_{0}^{1}
\sqrt{ g_{uv} \frac{d \tilde{\phi}^{u}}{d\varsigma} \frac{d \tilde{\phi}^{v}}{d\varsigma}+
	g_{qr}\frac{d \tilde{\varphi}^{q}}{d \varsigma} \frac{d \tilde{\varphi}^{r}}{d\varsigma} }
d\varsigma \label{eq16}
\end{equation}

We replace the $g_{uv}$ and $g_{qr}$ with their expressions from appendix \ref{secA}.
Then we use the following equalities:
\begin{equation}
\partial_{u} \rho_{\boldsymbol{\phi}}(\nu) \cdot \frac{d \phi^{u}}{d\varsigma}= \frac{\partial \rho_{\boldsymbol{\phi}}}{\partial \phi^{u}}(\nu)
\frac{d \phi^{u}}{d\varsigma}= \frac{d \rho_{\boldsymbol{\phi}}}{d\varsigma}(\nu) \nonumber \\
\end{equation}

\begin{equation}
\partial_{r} \psi_{\boldsymbol{\varphi}}(\nu) \cdot \frac{d \varphi^{r}}{d\varsigma}= \frac{\partial \psi_{\boldsymbol{\varphi}}}{\partial \varphi^{r}}(\nu)
\frac{d \varphi^{r}}{d\varsigma}= \frac{d \psi_{\boldsymbol{\varphi}}}{d\varsigma}(\nu) \nonumber \\
\end{equation}

We get:

\begin{equation}
d_{L2,\mathcal{B},\boldsymbol{\xi}}(\boldsymbol{\xi_{1}},\boldsymbol{\xi_{2}}) = \int_{0}^{1}
\sqrt{ \sum_{\nu \in \mathcal{B}} \dfrac{2}{\gamma_{0}(\nu)}  
	\left( \left( \frac{d \tilde{\rho}_{\boldsymbol{\phi}}}{d\varsigma}(\nu) \right)^2
	+  \left(\tilde{\rho}_{\boldsymbol{\phi}}(\nu)  \frac{d \tilde{\psi}_{\boldsymbol{\varphi}}}{d\varsigma}(\nu) \right)^2\right)  }
d\varsigma \label{eq17c}
\end{equation}

By definition, $\tilde{\rho}_{\boldsymbol{\xi}}=\tilde{\rho}_{\boldsymbol{\phi}}$ and $\tilde{\psi}_{\boldsymbol{\xi}}=\tilde{\psi}_{\boldsymbol{\varphi}}$.

Due to the derivative properties, we have $\frac{1}{\tilde{\rho}_{\boldsymbol{\xi}}(\nu)}\frac{d {\tilde{\rho}_{\boldsymbol{\xi}}}}{d\varsigma}(\nu) =\frac{d \ln{\tilde{\rho}_{\boldsymbol{\xi}}}}{d\varsigma}(\nu)$.

Using these equalities, equation \ref{eq17c} is easily transformed into equation \ref{eq17}.

\end{proof}

\begin{remark}
To compute the expression \ref{eq17} and give a closed-form expression of the Fisher-Rao distance, it is necessary to solve the equations \ref{eq14n}. This can only be done when the parametric expressions of magnitude $\rho_{\boldsymbol{\xi}}(\nu)$ and phase $\psi_{\boldsymbol{\xi}}(\nu)$ are expressed in closed forms, that is, when the model is particularised. This is done with the specific submanifold in the next section.
\end{remark}

\subsection{Geodesic equations on the manifold $L2(\mathcal{B},\alpha)$}\label{secD1}

Here we look at the geodesic equations on the submanifold containing finite-energy signals with a known magnitude spectrum and unknown attenuation and phase spectrum.

\begin{proposition}
    
 \label{prop3}
The (Fisher-)geodesic equations in the submanifold $L2(\mathcal{B},\alpha)$ reduce to the following equations:
\begin{equation}
\begin{cases}
\frac{d^{2} \alpha}{d\varsigma^{2}} 
-\alpha  \frac{1}{\omega_{0}} 
\sum_{\nu \in \mathcal{B}} \dfrac{2}{\gamma_{0}(\nu)} \left(\rho_{0}(\nu)\right)^2{\left( \frac{d \psi_{\boldsymbol{\varphi}}}{d\varsigma}(\nu)\right)^2 }
=  0  
\label{eq30a} \\
\forall \nu \in \mathcal{B} \ \ \ \ \ \alpha  \frac{d^{2} \psi_{\boldsymbol{\varphi}}}{d\varsigma^{2}}(\nu) 
+2 \frac{d\alpha}{d\varsigma} \frac{d \psi_{\boldsymbol{\varphi}}} {d\varsigma}(\nu) 
= 0

\end{cases}
\end{equation}
\end{proposition} 

\begin{proof} -\\
Based on equations \ref{eq19} and \ref{eq20}, we get:
\begin{align}
\frac{d \rho_{\boldsymbol{\phi}}}{d\varsigma}(\nu) & = \frac{d\alpha}{d\varsigma} \cdot \rho_{\boldsymbol{0}}(\nu) \nonumber \\
\frac{d^{2} \rho_{\boldsymbol{\phi}}}{d\varsigma^{2}}(\nu) & = \frac{d^{2} \alpha}{d\varsigma^{2}}
\cdot \rho_{\boldsymbol{0}}(\nu) \nonumber \\
\boldsymbol{\bigtriangledown}_{\boldsymbol{\phi}} \rho_{\boldsymbol{\phi}}(\nu) & = \rho_{\boldsymbol{0}}(\nu) \nonumber \\
\frac{d \psi_{\boldsymbol{\varphi}}}{d\varsigma}(\nu) & =\frac{d \breve{\psi}_{\boldsymbol{\varphi}}}{d\varsigma}(\nu)
=\sum_{r=2}^{N} \frac{d \varphi^r}{d\varsigma}  \cdot {(\nu)}^{r-2} \nonumber \\
\frac{d^{2} \psi_{\boldsymbol{\varphi}}}{d\varsigma^{2}}(\nu) & = \frac{d^{2} \breve{\psi}_{\boldsymbol{\varphi}}}{d\varsigma^{2}}(\nu)
=\sum_{r=2}^{N} \frac{d^2 \varphi^r}{d\varsigma^2}  \cdot {(\nu)}^{r-2}  \nonumber \\
\boldsymbol{\bigtriangledown}_{\boldsymbol{\varphi}} \psi_{\boldsymbol{\varphi}}(\nu) & =\boldsymbol{\bigtriangledown}_{\boldsymbol{\varphi}} \breve{\psi}_{\boldsymbol{\varphi}}(\nu)=
\begin{pmatrix}
1      \\
\nu^1 \\
... \\
\nu^{N-2} 
\end{pmatrix}
\label{50a55D}\nonumber
\end{align}
\\
We replace these values in equations \ref{eq44s}.

\end{proof}

\subsection{Fisher-Rao distance on the manifold $L2(\mathcal{B},\alpha)$}

As mentioned in section \ref{subsec44}, to obtain a closed-form expression of the Fisher-Rao distance, we must solve the geodesic equations \ref{eq30a}. We have the following solutions.
\begin{proposition}
    
 \label{prop3}
The solutions of the geodesic equations in the submanifold $L2(\mathcal{B},\alpha)$ are:
\begin{equation}
\begin{cases}
 \tilde{\alpha}(\varsigma) 
= \frac{\sqrt{k^2_1 (\varsigma+k_2)^2 +K}}{\sqrt{k_1}}  
\label{eq30u} \\
\forall \nu \in \mathcal{B} \ \ \ \ \ \frac{d \tilde{\psi}_{\boldsymbol{\varphi}}}{d\varsigma} 
= \frac{1}{\tilde{\alpha}^2} \cdot   \sqrt{K}\dfrac {\Delta \psi_{\boldsymbol{\varphi}}(\nu)}{\delta }

\end{cases}
\end{equation}
with:\\
    $\delta=\sqrt{\frac{1}{\omega_{0}} 
	\sum_{\nu \in \mathcal{B}} \dfrac {2} {\gamma_{0} (\nu) } \left(\rho_{0}(\nu)\right)^2{\left( \Delta \psi(\nu)\right)^2 }} $ \\
    $k_1= (\alpha_2)^2 + (\alpha_1)^2 - 2 \alpha_1 \alpha_2 \cos \delta$\\
    $k_2= \dfrac{- (\alpha_1)^2 + \alpha_1 \alpha_2 \cos \delta}{(\alpha_2)^2 + (\alpha_1)^2 - 2 \alpha_1 \alpha_2 \cos \delta}$\\
    $K= (\alpha_1)^2 (\alpha_2)^2 (\sin \delta)^2$
\end{proposition}

\begin{proof} -\\
Starting with equations \ref{eq30a}, we have:
\begin{equation}
\forall \nu \in \mathcal{B} \ \ \ \ \  \frac{1}{\frac{d \psi_{\boldsymbol{\varphi}}} {d\varsigma}(\nu) } \frac{d \frac{d \psi_{\boldsymbol{\varphi}}} {d\varsigma}}{d\varsigma}(\nu) 
+2 \frac{1} {\alpha}\frac{d\alpha}{d\varsigma}  
= 0
\label{50a55e}\nonumber
\end{equation}

This induces the following.
\begin{equation}
\forall \nu \in \mathcal{B} \ \ \ \exists c(\nu) \in \mathbb{R}\ \ :  \ \ \frac{d \psi_{\boldsymbol{\varphi}}} {d\varsigma}(\nu) = \frac{1} {\alpha^2} c(\nu) 
\label{50a551}\nonumber
\end{equation}
Using it in equation \ref{eq30a}, we obtain:
\begin{equation}
\frac{d^{2} \alpha}{d\varsigma^{2}} 
- \frac{1} {\alpha^3} \frac{1}{\omega_{0}} 
\sum_{\nu \in \mathcal{B}} \dfrac{2}{\gamma_{0}(\nu)} \left(\rho_{0}(\nu)\right)^2{\left(  c(\nu)\right)^2 }
=  0  
\label{50a552}\nonumber
\end{equation}
We set $K= \frac{1}{\omega_{0}} 
\sum_{\nu \in \mathcal{B}} \dfrac{2}{\gamma_{0}(\nu)} \left(\rho_{0}(\nu)\right)^2{\left( c(\nu)\right)^2 }$. \\
This leads to the following geodesic equations where the $c(\nu)$ are constant values (not depending on $\varsigma$).
\begin{equation}
\begin{cases}
\frac{d^{2} \alpha} {d\varsigma^{2}} 
= \frac{K}{\alpha^3}  
\label{eq30aa} \\
\forall \nu \in \mathcal{B} \ \ \ \ \ \frac{d \psi_{\boldsymbol{\varphi}}}{d\varsigma}(\nu) 
= \frac{1}{\alpha^2} \cdot  {c(\nu)} 

\end{cases}
\end{equation}

The values of all $c(\nu)$ and $K$ are determined by the boundary conditions.

See appendix \ref{secD}.
\end{proof}

\begin{remark}
As the difference between both signal phases $\Delta \psi(\nu)$ belongs to the interval $\left] -\pi, \pi\right]$ ($\Delta \psi(\nu)=\left\langle \psi_{\boldsymbol{\xi_2}}(\nu)-\psi_{\boldsymbol{\xi_1}}(\nu) \right\rangle_{\left] -\pi,\pi\right] }$),  we have to consider that $\delta=\sqrt{\frac{1}{\omega_{0}} 
	\sum_{\nu \in \mathcal{B}} \dfrac {2} {\gamma_{0} (\nu) } \left(\rho_{0}(\nu)\right)^2{\left( \Delta \psi(\nu)\right)^2 }} $ belongs to the segment $\delta \in \left[0 , \pi\right]$. This is consistent with the fact that $\delta$ is also a difference of two $\arctan$ functions (see appendix \ref{secD}).
\end{remark}

\begin{theorem} \label{th4}
The Fisher-Rao distance on the manifold $L2(\mathcal{B},\alpha)$ is given by:
\begin{equation}
d_{L2,\mathcal{B},\alpha}(\boldsymbol{\xi_{1}},\boldsymbol{\xi_{2}}) 
 = \sqrt{\omega_{0}} \cdot 
\sqrt{(\alpha_2)^2 + (\alpha_1)^2 - 2 \alpha_1 \alpha_2 \cos \left( \sqrt{\frac{1}{\omega_{0}} 
	\sum_{\nu \in \mathcal{B} } \dfrac {2} {\gamma_{0} (\nu) } \left(\rho_{0}(\nu)\right)^2 {\left( \Delta \psi(\nu)\right)^2 }} \right) } 
\label{eq23cc}
\end{equation}
\end{theorem}
\begin{proof}- \\

We have the following (see appendix \ref{secD}).
\begin{align}
    &\tilde{\alpha}  =  \frac{\sqrt{k^2_1 (\varsigma+k_2)^2 +K}}{\sqrt{k_1}} \nonumber \\
    &\frac{d\tilde{\alpha}}{d\varsigma}  = \frac{k_1 \sqrt{k_1} (k_2+\varsigma)}{ \sqrt{k^2_1 (\varsigma+k_2)^2 +K} } \nonumber \\
    &\rho_{\boldsymbol{\phi}}  = \alpha  \cdot \rho_{\boldsymbol{0}}(\nu) \nonumber \\
   & \frac{d \rho_{\boldsymbol{\phi}}}{d\varsigma}(\nu)  = \frac{d\alpha}{d\varsigma} \cdot \rho_{\boldsymbol{0}}(\nu) \nonumber \\
    &\rho_{\boldsymbol{\phi}}(\nu) \frac{d \psi_{\boldsymbol{\varphi}}}{d\varsigma}(\nu) = \frac{1}{\alpha^2} \cdot  {c(\nu)} 
\cdot \rho_{\boldsymbol{\phi}}(\nu) \nonumber 
\end{align}

Replacing the values in equation \ref{eq17c}, we get:
\begin{equation}
d_{L2,\mathcal{B},\alpha}(\boldsymbol{\xi_{1}},\boldsymbol{\xi_{2}}) = \int_{0}^{1}
\sqrt{ \left( \frac{d\tilde{\alpha}}{d\varsigma}\right)^2 \omega_{0}
	+ \frac{1}{\tilde{\alpha}^2}  
\sum_{\nu \in \mathcal{B}} \dfrac{2}{\gamma_{0}(\nu)} \left(\rho_{0}(\nu)\right)^2{\left( c(\nu)\right)^2 }	  }
d\varsigma \label{eq84D}
\end{equation}
which becomes:
\begin{equation}
d_{L2, \mathcal{B},\alpha}(\boldsymbol{\xi_{1}},\boldsymbol{\xi_{2}}) 
= \sqrt{\omega_{0}} \int_{0}^{1}
\sqrt{ \left( \frac{d\tilde{\alpha}}{d\varsigma}\right)^2 
	+\frac{K}{\tilde{\alpha}^2}  }
d\varsigma \label{eq85D}
\end{equation}
and finally:
\begin{equation}
d_{L2, \mathcal{B},\alpha} (\boldsymbol{\xi_{1}},\boldsymbol{\xi_{2}}) 
= \sqrt{\omega_{0} k_1} 
\label{eq86D}
\end{equation}
Which is:
\begin{align}
& d_{L2, \mathcal{B},\alpha}(\boldsymbol{\xi_{1}},\boldsymbol{\xi_{2}}) \notag \\
& = \sqrt{\omega_{0}} \cdot 
\sqrt{(\alpha_2)^2 + (\alpha_1)^2 - 2 \alpha_1 \alpha_2 \cos \left( \sqrt{\frac{1}{\omega_{0}} 
	\sum_{\nu \in \mathcal{B}} \dfrac{2}{\gamma_{0}(\nu)} \left(\rho_{0}(\nu)\right)^2{\left( \Delta \psi(\nu)\right)^2  }}\right) } 
\label{eq87D}
\end{align}
\\
\end{proof}

\section{Evaluation of the Fisher-Rao distances} \label{sec5}

\subsection{Comparison of the Fisher-Rao distances on the $L2(\mathcal{B})$ manifold and on any $L2(\mathcal{B},\boldsymbol{\xi})$ submanifold}

The $N$-dimensional submanifolds $L2(\mathcal{B},\boldsymbol{\xi})$ are included in the $2N_{\mathcal{B}}$-dimensional manifold $L2(\mathcal{B})$ composed by multivariate normal distributions with a common diagonal covariance matrix. As a consequence, we expect that the Fisher-Rao distance $d_{L2,\mathcal{B},\boldsymbol{\xi}}\left(\boldsymbol{\xi_{1}},\boldsymbol{\xi_{2}} \right) $ measured on any submanifold $L2(\mathcal{B},\boldsymbol{\xi})$ is greater than or equal to the Fisher-Rao distance measured on the manifold $L2(\mathcal{B})$:
\begin{equation}
d_{L2,\mathcal{B},\boldsymbol{\xi}}\left(\boldsymbol{\xi_{1}},\boldsymbol{\xi_{2}} \right) \geq
d_{L2,\mathcal{B}} \left(\boldsymbol{\xi_{1}},\boldsymbol{\xi_{2}}  \right)  
\label{eq18}
\end{equation}
When both quantities are always equal, the submanifold $L2(\mathcal{B},\boldsymbol{\xi})$ is said to be fully geodesic on the manifold $L2(\mathcal{B})$.

\begin{remark}
To measure a Fisher-Rao distance between two points on a submanifold $L2(\mathcal{B},\boldsymbol{\xi})$, both points must belong to the submanifold. When we want to compare the distance measured on a submanifold $L2(\mathcal{B},\boldsymbol{\xi})$ with the distance measured on the manifold $L2(\mathcal{B})$, both points must belong to the submanifold $L2(\mathcal{B},\boldsymbol{\xi})$. In what follows, we compare the distances on the submanifold $L2(\mathcal{B},\boldsymbol{\alpha})$ and on the manifold $L2(\mathcal{B})$. As a consequence, the points $\boldsymbol{\xi_{1}}$ and $\boldsymbol{\xi_{2}}$ belong to the submanifold $L2(\mathcal{B},\boldsymbol{\alpha})$.
\end{remark}

\subsection{Asymptotic behaviour of the Fisher-Rao distances}

By rewriting the Fisher-Rao distances with the ratio of magnitudes $\gamma=\frac{\alpha_2}{\alpha_1}$ and the signal-to-noise ratio on the signal at $\boldsymbol{\xi_{1}}$, that is $SNR_1=\omega_0 \left( \alpha_1\right)^2$, we derive the expressions of the Fisher-Rao distances on the manifold $L2(\mathcal{B})$ of all finite-energy signals and on the submanifold $L2(\mathcal{B},\alpha)$ of finite-energy signals with a known magnitude spectrum $\rho_0\left(\nu\right)$.
From equations \ref{eq24} and \ref{eq23cc}, we see that the Fisher-Rao distances on the manifold $L2(\mathcal{B})$ and on the manifold $L2(\mathcal{B},\alpha)$ are respectively:

\begin{equation}
d_{L2,\mathcal{B}} \left(\boldsymbol{\xi_{1}},\boldsymbol{\xi_{2}}  \right) 
=\sqrt{SNR_1} \cdot \sqrt{  \left( \gamma \right)^2 
	+ 1- 2  \gamma \cdot {\frac{1}{\omega_{0}} \sum_{\nu \in \mathcal{B}} \dfrac {2} {\gamma_{0}(\nu) } \left(\rho_{0}(\nu) \right)^2 \cos\left(\Delta \psi (\nu) \right) }}
\label{eq25}
\end{equation}

\begin{equation}
d_{L2,\mathcal{B},\alpha}\left(\boldsymbol{\xi_{1}},\boldsymbol{\xi_{2}}  \right)  
= \sqrt{SNR_1} \cdot 
\sqrt{(\gamma)^2 + 1- 2 \gamma \cos \left( \sqrt{\frac{1}{\omega_{0}} 
	\sum_{\nu \in \mathcal{B}} \dfrac{2}{\gamma_{0}(\nu)} \left(\rho_{0}(\nu) \right)^2 {\left( \Delta \psi(\nu)\right)^2 }} \right) } 
\label{eq26}
\end{equation}
The Fisher-Rao distances are proportional to the square root of the signal-to-noise ratio. Therefore, we can study the geometry for a reference level $SNR_1=1$ (0 dB).\\

Additionally, the distances depend on the ratio $\gamma$ between the attenuations $\alpha_1$ and $\alpha_2$.

\begin{proposition}
When the phase differences $\Delta \psi(\nu)$ are small, both distances are equivalent:
\begin{equation}
d_{L2,\mathcal{B},\alpha}\left(\boldsymbol{\xi_{1}},\boldsymbol{\xi_{2}}  \right)  
\underset{\Delta \psi(\nu) \to 0}{\sim} d_{L2,\mathcal{B}} \left(\boldsymbol{\xi_{1}},\boldsymbol{\xi_{2}}  \right)
\underset{\Delta \psi(\nu) \to 0}{\sim} \sqrt{SNR_1} \cdot 
\sqrt{(\gamma- 1)^2+ \gamma   \frac{1}{\omega_{0}} 
	\sum_{\nu \in \mathcal{B}} \dfrac{2}{\gamma_{0}(\nu)} \left(\rho_{0}(\nu) \right)^2 \left( \Delta \psi(\nu)\right)^2 }  
\label{eq27}
\end{equation}
\end{proposition}

\begin{proof}
  Use a Taylor expansion of $\cos$ in equations \ref{eq25} and \ref{eq26}.
\end{proof}

\begin{proposition}
When the phases differences are constant across the bandwidth ($\Delta \psi(\nu)=\Delta \psi_0$) both distances are equal:
\begin{equation}
d_{L2,\mathcal{B},\alpha}\left(\boldsymbol{\xi_{1}},\boldsymbol{\xi_{2}}  \right)  
\underset{\Delta \psi(\nu) = \Delta \psi_0}{=} d_{L2,\mathcal{B}} \left(\boldsymbol{\xi_{1}},\boldsymbol{\xi_{2}}  \right) 
\underset{\Delta \psi(\nu) = \Delta \psi_0}{=} \sqrt{SNR_1} \cdot 
\sqrt{(\gamma)^2 + 1- 2 \gamma \cos \left(  \Delta \psi_0 \right) }   
\label{eq28}
\end{equation}  
\end{proposition}

  \begin{proof}
Replace $\Delta \psi(\nu)$ by $\Delta \psi_0$ in equations \ref{eq25} and \ref{eq26}.
\end{proof}

  \begin{proposition}
When the phases differences tend to a constant value across the bandwidth ($\Delta \psi(\nu)\to\Delta \psi_0$) both distances tend to the same value:

\begin{equation}
\lim_{\Delta \psi(\nu) \to \Delta \psi_0} d_{L2,\mathcal{B},\alpha}\left(\boldsymbol{\xi_{1}},\boldsymbol{\xi_{2}}  \right)  
= \lim_{\Delta \psi(\nu) \to \Delta \psi_0} d_{L2,\mathcal{B}} \left(\boldsymbol{\xi_{1}},\boldsymbol{\xi_{2}}  \right) 
= \sqrt{SNR_1} \cdot 
\sqrt{(\gamma)^2 + 1- 2 \gamma \cos \left(  \Delta \psi_0 \right) }   
\label{eq29}
\end{equation}
\end{proposition}
\begin{proof}
Use $\Delta \psi(\nu) \to \Delta \psi_0$ in equations \ref{eq25} and \ref{eq26}.
\end{proof}

\begin{proposition}
For large phase variations, we have:
\begin{equation}
 d_{L2,\mathcal{B}} \left(\boldsymbol{\xi_{1}},\boldsymbol{\xi_{2}}  \right) 
\approx \sqrt{SNR_1} \cdot \sqrt{  \left( \gamma \right)^2 + 1}
\label{eq30}
\end{equation}
and:
\begin{equation}
d_{L2,\mathcal{B},\alpha}\left(\boldsymbol{\xi_{1}},\boldsymbol{\xi_{2}}  \right)  
\approx \sqrt{SNR_1} \cdot 
\sqrt{(\gamma)^2 + 1- 2 \gamma \cos \left( \frac{\pi}{\sqrt{3}}\right) }
\label{eq31}
\end{equation}
\end{proposition}
  \begin{proof}- \\
When the phase differences $\Delta \psi(\nu)$ vary quickly across the bandwidth, we can assume that:
\begin{equation}
\frac{1}{\omega_{0}} \sum_{\nu \in \mathcal{B}} \dfrac {2} {\gamma_{0} (\nu) } \left(\rho_{0}(\nu)\right)^2 \cos\left(\Delta \psi (\nu) \right) \approx E\left( \cos(\theta)\right)
\label{eq32}
\end{equation}
and
\begin{equation}
\cos \left( \sqrt{\frac{1}{\omega_{0}} 
	\sum_{\nu \in \mathcal{B}} \dfrac{2}{\gamma_{0}(\nu)} \left(\rho_{0}(\nu) \right)^2 {\left( \Delta \psi(\nu)\right)^2 }} \right) \approx \cos \left( \sqrt{E\left( \left( \theta \right)^2\right) } \right) \label{eq33}
\end{equation}
where $\theta$ is a random angle between $-\pi$ and $\pi$. \\

We assume a constant probability density function and get $E\left( \cos(\theta)\right)=0$ and $\cos \left( \sqrt{E\left( \left( \theta \right)^2\right) } \right)=\cos \left( \frac{\pi}{\sqrt{3}}\right)$. \\
\end{proof}

\subsection{Comparison of the Fisher-Rao distances with numerical applications}

If we assume that the signal-to-noise ratio $\dfrac{2}{\gamma_{0}(\nu)} \left(\rho_{0}(\nu) \right)^2$ is constant within the bandwidth, from equations \ref{eq25} and \ref{eq26}, the ratio between the distances becomes:
\begin{equation}
\dfrac{d_{L2,\mathcal{B},\alpha}\left(\boldsymbol{\xi_{1}},\boldsymbol{\xi_{2}}  \right) }
{d_{L2,\mathcal{B}}\left(\boldsymbol{\xi_{1}},\boldsymbol{\xi_{2}}  \right) } 
= \sqrt{ \dfrac{	(\gamma)^2 + 1 - 2 \gamma \cos \left( 
	\sqrt{ \frac{1} {N_\mathcal{B}} \sum_{\nu\in \mathcal{B}} \left(\Delta \psi(\nu) \right)^2 } \right) }	
	 { 	(\gamma)^2+ 1- 2  \gamma \frac{1}{N_{\mathcal{B}}} \sum_{\nu\in \mathcal{B} }  \cos \left( \Delta \psi(\nu) \right)	} 
	 }
\label{eq35}
\end{equation}

For numerical application, we specify the model associated with the phase.\\

The signal is supposed to be time-delayed (with a phase varying linearly with the frequency), that is:
\begin{equation}
\breve{\psi}_{\boldsymbol{\varphi}} (\nu)=\psi_0 - 2 \pi \tau \nu
\label{eq36}
\end{equation}
In addition, we assume that $\check{B}=\left[\nu_0-\frac{B}{2},\nu_0+\frac{B}{2} \right]$ and from equation \ref{eq35} we get:

\begin{equation}
\dfrac{d_{L2,\mathcal{B},\alpha}\left(\boldsymbol{\xi_{1}},\boldsymbol{\xi_{2}}  \right) }
{d_{L2,\mathcal{B}}\left(\boldsymbol{\xi_{1}},\boldsymbol{\xi_{2}}  \right) } 
\approx \sqrt{ \dfrac{	(\gamma)^2 + 1 - 2 \gamma \cos \left( 
	\sqrt{ \frac{1} {N_\mathcal{B}} \sum_{\nu \in \mathcal{B}} \left(\left\langle \Delta \psi_0 - 2 \pi \nu \Delta \tau \right\rangle_{\left] -\pi,\pi\right] } \right)^2 } \right)  }	
	{ 	(\gamma)^2+ 1- 2  \gamma  \dfrac{\sin\left( \pi B \Delta \tau\right) }{\pi B \Delta \tau} 
	\cos \left( \Delta \psi_0 - 2 \pi \nu_0 \Delta \tau  \right)	} 
}
\label{eq37}
\end{equation}
Where:\\
$\Delta \psi_0 = \psi_0\left(\boldsymbol{\xi_2} \right)- \psi_0\left(\boldsymbol{\xi_1} \right)$ \\
$\Delta \tau=\tau\left(\boldsymbol{\xi_2} \right)- \tau \left(\boldsymbol{\xi_1} \right)$\\
$\left\langle \Delta \psi_0 - 2 \pi \nu \Delta \tau \right\rangle_{\left] -\pi,\pi\right] } $ is the angle restricted to the interval between $-\pi$ and $\pi$.\\

From equation \ref{eq18}, we expect the following property:\\

\begin{equation}
\dfrac{d_{L2,\mathcal{B},\alpha}\left(\boldsymbol{\xi_{1}},\boldsymbol{\xi_{2}}  \right)}
{d_{L2,\mathcal{B}} \left(\boldsymbol{\xi_{1}},\boldsymbol{\xi_{2}}  \right)}  
\geq  1 
\label{eq34}
\end{equation}

This is confirmed in the following figures.

\subsection{Figures} \label{sec55}

We use the expressions of the distances from equation \ref{eq37}, with the reference signal-to-noise ratio $ SNR_1=1 $.
Some of the parameters are fixed: $N_\mathcal{B}=1000$, $\nu_0=0.25$.
$\Delta \tau$ varies, and we draw the figures as a function of $B \Delta \tau$ (x-coordinate).

For the other parameters, we study different cases:
\begin{itemize}
\item $B=0.5$, $\Delta \psi_0 =0$, $\gamma=1$\\

This case focusses on wideband signals with the same attenuation and phase offset, but different time delays. As seen in equations \ref{eq30} and \ref{eq31}, the ratio tends to a limit $\sqrt{1-\cos\left( \frac{\pi}{\sqrt{3}}\right)} \approx 1.11$ for large delays.
For small delays, the distance values remain the same. \\
\begin{figure}[H]
\begin{minipage}[H]{0.6\textwidth}
\includegraphics[width=\textwidth]{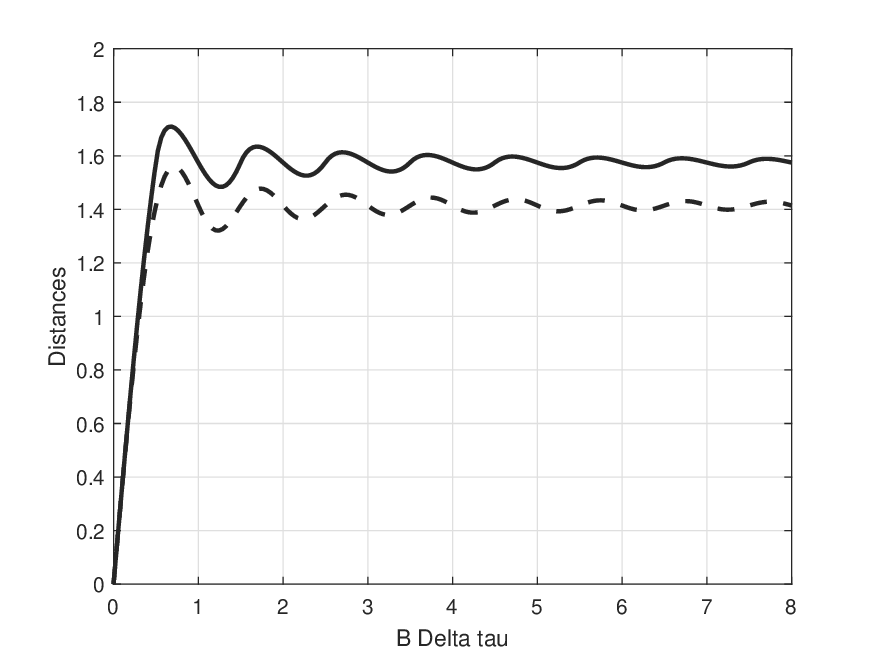}
\caption{Comparison of the two distances as functions of $B \Delta \tau $. The lowest distance is $d_{L2,\mathcal{B}}\left(\boldsymbol{\xi_{1}},\boldsymbol{\xi_{2}}  \right)$ with the dashed line. $B=0.5$, $\Delta \psi_0 =0$, $\gamma=1$.} \label{fig1a}
\end{minipage}
\hfill 
\hspace{0.05\textwidth}
\begin{minipage}[H]{0.6\textwidth}
\includegraphics[width=\textwidth]{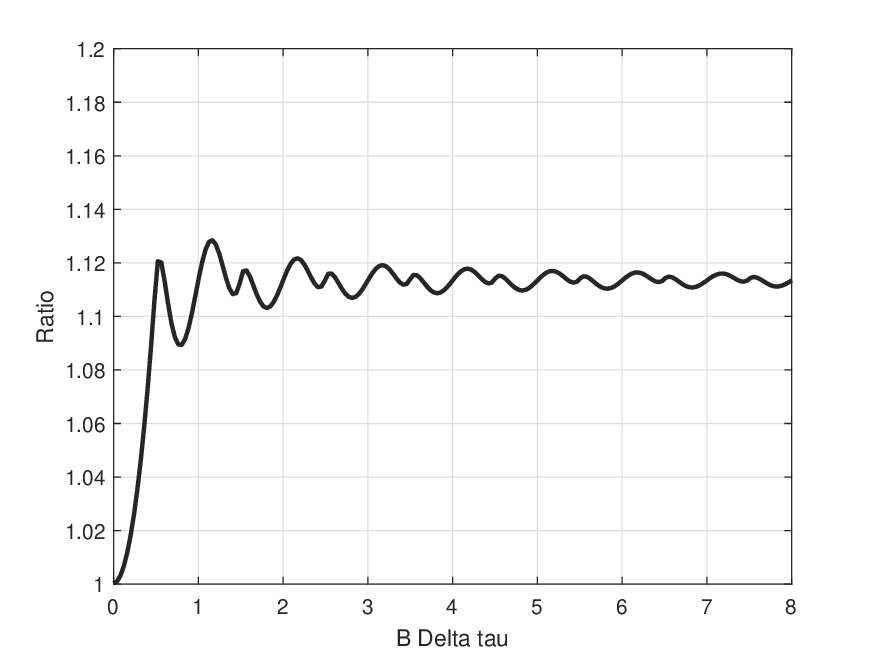}
\caption{Ratio of the two distances as a function of $B \Delta \tau $. $B=0.5$, $\Delta \psi_0 =0$, $\gamma=1$. } \label{fig2a}
\end{minipage}
\end{figure}

\item $B=0.5$, $\Delta \psi_0 =\pi/2$, $\gamma=1$\\

This case focusses on wideband signals with the same attenuation, but different time delays and phase offsets. For large delays, the ratio tends to $1.11$ as in the previous case.
For small delays, the distance values remain the same. Note that there is a decrease for some delays, due to the difference in phase offsets. In this case, the distances do not cancel out because the mean value of the phase differences across the bandwidth never cancels. The minimum distance is not zero. \\
\begin{figure}[H]
\begin{minipage}[H]{0.6\textwidth}
\includegraphics[width=\textwidth]{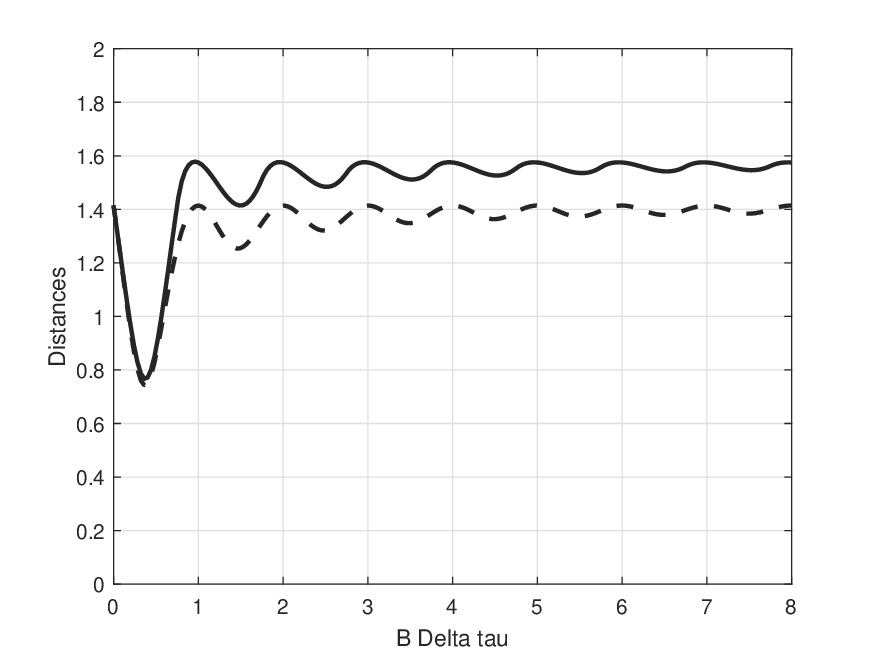}
\caption{Comparison of the two distances as functions of $B \Delta \tau $. The lowest distance is $d_{L2,\mathcal{B}}\left(\boldsymbol{\xi_{1}},\boldsymbol{\xi_{2}}  \right)$ with the dashed line. $B=0.5$, $\Delta \psi_0 =\pi/2$, $\gamma=1$. } \label{fig3a}
\end{minipage}
\hfill 
\hspace{0.05\textwidth}
\begin{minipage}[H]{0.6\textwidth}
\includegraphics[width=\textwidth]{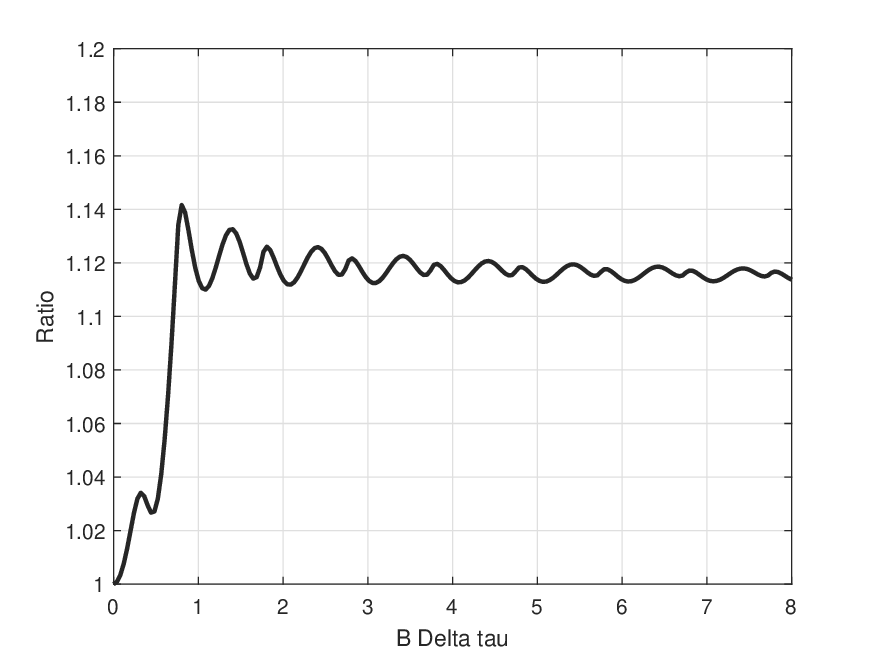}
\caption{Ratio of the two distances as a function of $B \Delta \tau $. $B=0.5$, $\Delta \psi_0 =\pi/2$, $\gamma=1$. } \label{fig4a}
\end{minipage}
\end{figure}

\item $B=0.5$, $\Delta \psi_0 =0$, $\gamma=10$\\

This case focusses on wideband signals with the same phase offset but different attenuations and  time delays. As seen in equations \ref{eq30} and \ref{eq31}, the ratio tends to a limit $\sqrt{1-\frac{20}{101}\cos\left( \frac{\pi}{\sqrt{3}}\right)} \approx 1.02$ for large delays.
For small delays, the distance values remain the same.\\
\begin{figure}[H]
\begin{minipage}[H]{0.6\textwidth}
\includegraphics[width=\textwidth]{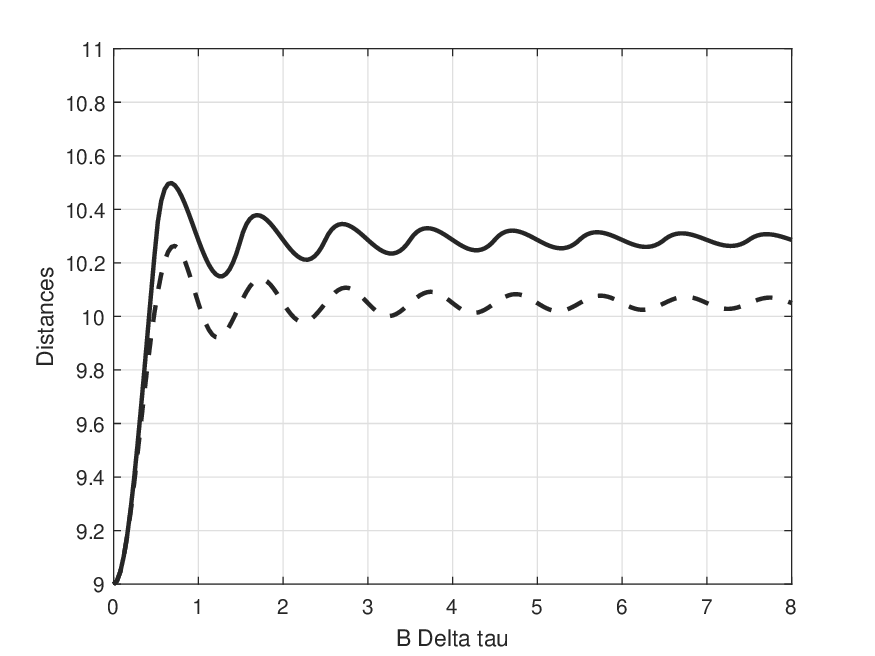}
\caption{Comparison of the two distances as functions of $B \Delta \tau $. The lowest distance is $d_{L2,\mathcal{B}}\left(\boldsymbol{\xi_{1}},\boldsymbol{\xi_{2}}  \right)$ with the dashed line. $B=0.5$, $\Delta \psi_0 =0$, $\gamma=10$. } \label{fig5a}
\end{minipage}
\hfill 
\hspace{0.05\textwidth}
\begin{minipage}[H]{0.6\textwidth}
\includegraphics[width=\textwidth]{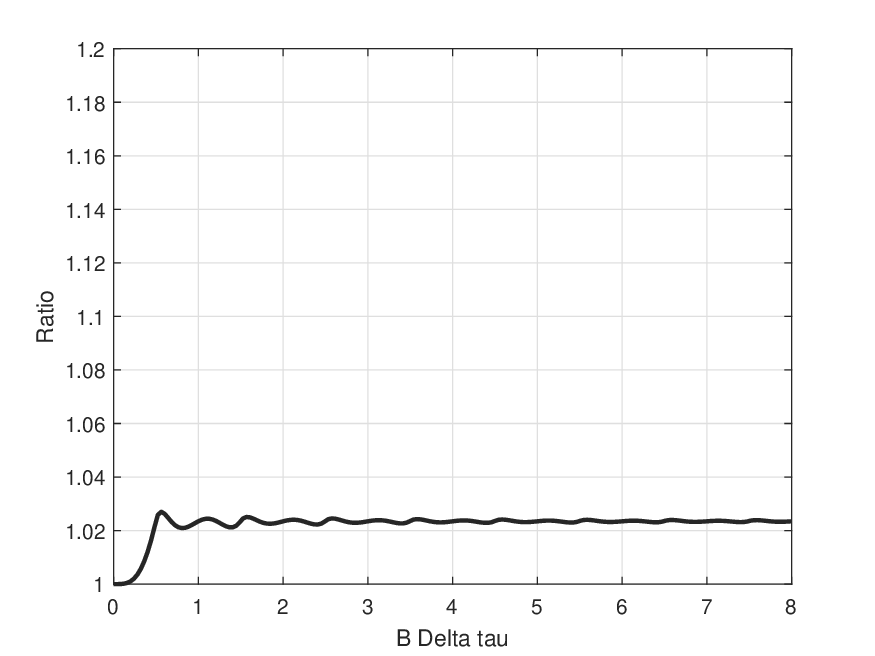}
\caption{Ratio of the two distances as a function of $B \Delta \tau $. $B=0.5$, $\Delta \psi_0 =0$, $\gamma=10$. } \label{fig6a}
\end{minipage}
\end{figure}

\item $B=0.25$, $\Delta \psi_0 =0$, $\gamma=1$\\
This case focusses on lower band signals with the same attenuation and phase offset, but different time delays. The ratio again tends to $1.11$ for large delays.
For small delays, the distance values remain the same.\\
\begin{figure}[H]
\begin{minipage}[H]{0.6\textwidth}
\includegraphics[width=\textwidth]{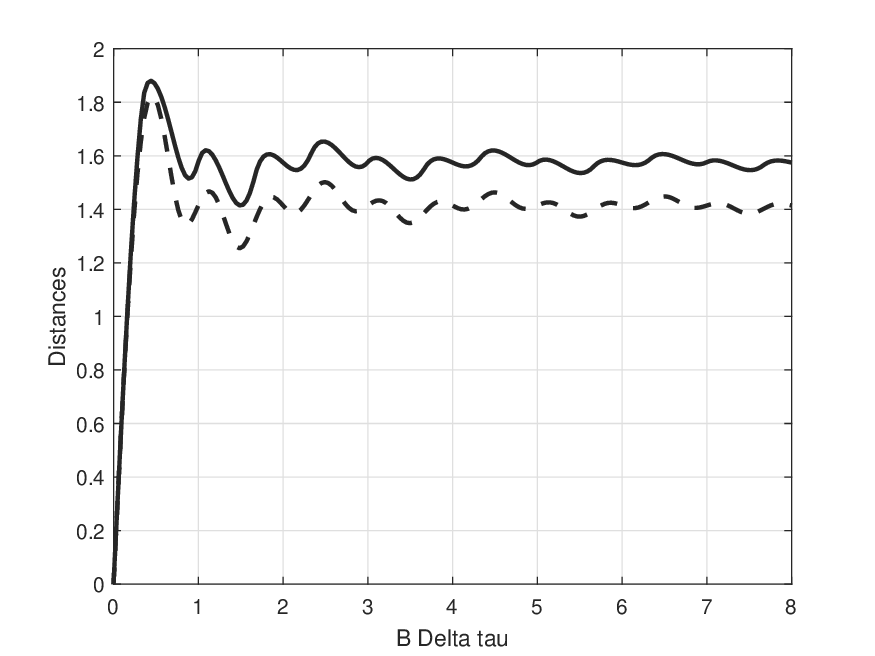}
\caption{Comparison of the two distances as functions of $B \Delta \tau $. The lowest distance is $d_{L2,\mathcal{B}}\left(\boldsymbol{\xi_{1}},\boldsymbol{\xi_{2}}  \right)$ with the dashed line. $B=0.25$, $\Delta \psi_0 =0$, $\gamma=1$. } \label{fig7a}
\end{minipage}
\hfill 
\hspace{0.05\textwidth}
\begin{minipage}[H]{0.6\textwidth}
\includegraphics[width=\textwidth]{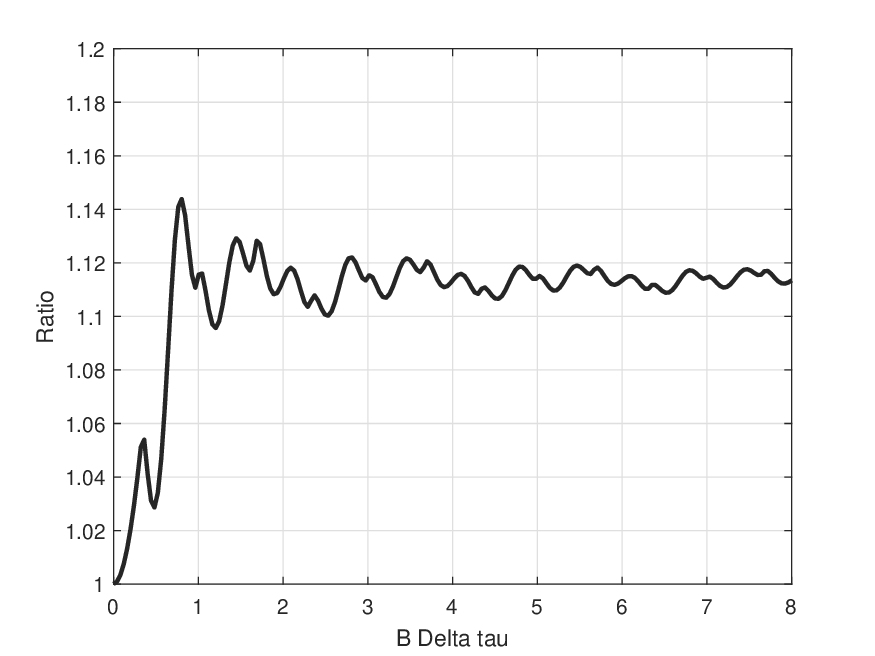}
\caption{Ratio of the two distances as a function of $B \Delta \tau $. $B=0.25$, $\Delta \psi_0 =0$, $\gamma=1$. } \label{fig8a}
\end{minipage}
\end{figure}

\end{itemize}

%\begin{figure} 
%\includegraphics[width=10 cm]{Distances.jpg}\hfill 
%\includegraphics[width=10 cm]{Courbe.jpg} 
%\caption{Titre commun}\label{fig:somefiglabel} 
%\end{figure} 

\section{Conclusion}\label{sec13}

In this work, we focussed on the parametric modelling of finite-energy signals in the presence of additive noise. We assumed a scenario where the noise is circularly symmetric zero mean complex Gaussian after a Fourier transform and is independent across frequencies. The observation vector consists of the Fourier transform outputs within a specified observation bandwidth. The noise spectral density is known, as is commonly the case in many applications. The observation vector is random. It depends on random noise samples and on deterministic signal parameters. To assess the difficulty in distinguishing between two signals based on the statistical distributions of their observations, we considered the information geometry approach. We modelled the signal observations as points on a statistical manifold and calculated the Fisher-Rao distance between the two corresponding points on the manifold.

To achieve this, we used the block diagonal structure of the Fisher matrix of the model. This matrix induces a Riemannian metric on the statistical manifold defined by the signal parameters. We estimated the Christoffel symbols of the Levi-Civita connexion and the tensorial equations of the geodesics on the manifolds. Based on the finite-energy signal model, we demonstrated that for any parametric description of finite-energy signals, geodesics are described by two combined partial differential equations involving phase spectrum and magnitude spectrum and their gradients (LDG theorem). This LDG theorem expresses the linear dependence of the phase and magnitude gradients relative to their respective parameters.

We have examined two statistical situations corresponding to two statistical manifolds from the geometric science of information perspective. The first situation corresponds to the manifold of all possible finite-energy signals observed in the given bandwidth, which we referred to as the manifold $L2(\mathcal{B})$. We demonstrated that this manifold corresponds to a multivariable normal distribution with the same covariance matrix but different mean values.
The second situation corresponds to the estimation of finite-energy signals when the magnitude spectrum $\rho_0(\nu)$ is known, except for a global attenuation coefficient $\alpha$. This manifold was called the manifold $L2(\mathcal{B},\alpha)$.

Finding closed-form expressions for the Fisher-Rao distance is generally a non-trivial task. We obtained a closed-form expression for the Fisher-Rao distance on the global manifold $L2(\mathcal{B})$ that includes all finite-energy signals. For this case, we deduced from the literature that the Fisher-Rao distance is equivalent to the Mahalanobis distance. We also obtained a closed-form expression for the Fisher-Rao distance on the sub-manifold $L2(\mathcal{B},\alpha)$ that includes the finite-energy signal with a known magnitude spectrum, an unknown phase spectrum, and an unknown attenuation. The work detailed the solutions of the geodesic equations. We demonstrated that the sub-manifold $L2(\mathcal{B},\alpha)$ is not fully geodesic on the entire manifold $L2(\mathcal{B})$.

The analysis of the Fisher-Rao distance expressions revealed noteworthy results regarding the estimation problem. The first simple property is that the Fisher-Rao distance is proportional to the signal-to-noise ratio at the chosen reference signal. This means that the geometric structure of the manifold is controlled by a homothetical transformation, scaled by the signal-to-noise ratio.

Another property is that the Fisher-Rao distance between two signals varies as a function of the ratios of their spectral magnitudes. 

Additionally, the difference in phase spectrums of two signals impacts their Fisher-Rao distance. When the difference in phase spectrums is small or does not vary throughout the bandwidth, the distance values remain the same on the submanifold  $L2(\mathcal{B},\alpha)$ and on the manifold $L2(\mathcal{B})$. This indicates that knowing the magnitude spectrum has a limited impact on the phase parameter estimation. For a signal with an unknown constant phase, knowing the signal magnitude spectrum does not increase the Fisher-Rao distance, that is, it does not make the phase estimation easier. Conversely, when the difference in the phase spectrum varies significantly throughout the bandwidth, the knowledge of the magnitude spectrum provides an advantage for signal parameter estimation. The constraint provided by knowing the magnitude spectrum increases the Fisher-Rao distance. For signals with the same energy, the ratio is $\sqrt{1-\cos\left( \frac{\pi}{\sqrt{3}}\right)} \approx 1.11$. This scenario occurs, for instance, when wideband signals are separated by a sufficient time delay.

Although we obtained simplified equations for geodesics, finding their general solution for any type of constraint in signal estimation remains an open problem. For instance, we did not consider specific constraints as minimum phase or specific parametric expression inducing dependence between magnitude and phase. In the same way, other cases of controlled signal magnitude spectrum should be investigated. Nevertheless, we anticipate that the Fisher-Rao distance between finite energy signals will help the characterisation of signal databases. Additionally, geodesic equations and the LDG theorem may provide valuable insights for some estimation techniques, such as analytic-informed neural networks. For example, equations \ref{eq14n} or \ref{eq44s} could be used in the forward/backward estimation process of unknown coefficients of a deep neural network.

\backmatter

\bmhead{Acknowledgements}
The author thanks Dr Mona Florin for her positive conversations and the anonymous reviewers for their excellent feedback. These helped me a lot in clarifying the presentation.

\section*{Declarations}
\begin{itemize}
\item Funding \\
No funding was received for the conduct of this study.
\item Conflict of interest / Competing interests \\
The author has no conflicts of interest to declare that are relevant to the content of this article.

\end{itemize}

\begin{appendices}

\section{Fisher metric}\label{secA}

From equations $\ref{eq01}$ and $\ref{eq07} $, we deduce:

\begin{align}
g_{ij} &= \sum_{\nu}\dfrac{1}{\gamma_{0}^2(\nu)} E_{\boldsymbol{\xi}}\left[ \frac{\partial}{\partial \xi^i} { \left|  x(\nu)-s_{\boldsymbol{\xi}} (\nu)   \right|^2   }
\frac{\partial}{\partial \xi^j} { \left|  x(\nu)-s_{\boldsymbol{\xi}} (\nu)   \right|^2   } \right] \nonumber \\
&= \sum_{\nu}\dfrac{1}{\gamma_{0}^2(\nu)} E_{\boldsymbol{\xi}}\left[ \frac{\partial}{\partial \xi^i} { \left( \left|  x(\nu) \right|^2-2  Re \left\lbrace x(\nu)^* s_{\boldsymbol{\xi}} (\nu) \right\rbrace + \left| s_{\boldsymbol{\xi}} (\nu)   \right|^2 \right)  }
\frac{\partial}{\partial \xi^j} { \left( \left|  x(\nu) \right|^2-2  Re \left\lbrace x(\nu)^* s_{\boldsymbol{\xi}} (\nu) \right\rbrace + \left| s_{\boldsymbol{\xi}} (\nu)   \right|^2 \right) } \right] \nonumber \\
&= \sum_{\nu}\dfrac{1}{\gamma_{0}^2(\nu)} E_{\boldsymbol{\xi}}\left[  { \left( -2  Re \left\lbrace x(\nu)^*\frac{\partial}{\partial \xi^i} s_{\boldsymbol{\xi}} (\nu) \right\rbrace + \frac{\partial}{\partial \xi^i}\left| s_{\boldsymbol{\xi}} (\nu)   \right|^2 \right)  }
{ \left( -2  Re \left\lbrace x(\nu)^* \frac{\partial}{\partial \xi^j} s_{\boldsymbol{\xi}} (\nu) \right\rbrace + \frac{\partial}{\partial \xi^j}  \left| s_{\boldsymbol{\xi}} (\nu)   \right|^2 \right) } \right] \nonumber \\
&= \sum_{\nu}\dfrac{1}{\gamma_{0}^2(\nu)} E_{\boldsymbol{\xi}}\left[  { \left( -2  Re \left\lbrace x(\nu)^*\frac{\partial}{\partial \xi^i} s_{\boldsymbol{\xi}} (\nu) \right\rbrace \right)  }
{ \left( -2  Re \left\lbrace x(\nu)^* \frac{\partial}{\partial \xi^j} s_{\boldsymbol{\xi}} (\nu) \right\rbrace  \right) } \right] \nonumber \\
&+ \sum_{\nu}\dfrac{1}{\gamma_{0}^2(\nu)} E_{\boldsymbol{\xi}}\left[  { \left( -2  Re \left\lbrace x(\nu)^*\frac{\partial}{\partial \xi^i} s_{\boldsymbol{\xi}} (\nu) \right\rbrace \right)  }
{ \left(  \frac{\partial}{\partial \xi^j}  \left| s_{\boldsymbol{\xi}} (\nu)   \right|^2 \right) } \right] \nonumber \\
&+\sum_{\nu}\dfrac{1}{\gamma_{0}^2(\nu)} E_{\boldsymbol{\xi}}\left[  { \left(  \frac{\partial}{\partial \xi^i}\left| s_{\boldsymbol{\xi}} (\nu)   \right|^2 \right)  }
{ \left( -2  Re \left\lbrace x(\nu)^* \frac{\partial}{\partial \xi^j} s_{\boldsymbol{\xi}} (\nu) \right\rbrace  \right) } \right] \nonumber \\
&+\sum_{\nu}\dfrac{1}{\gamma_{0}^2(\nu)} E_{\boldsymbol{\xi}}\left[  { \left(  \frac{\partial}{\partial \xi^i}\left| s_{\boldsymbol{\xi}} (\nu)   \right|^2 \right)  }
{ \left(  \frac{\partial}{\partial \xi^j}  \left| s_{\boldsymbol{\xi}} (\nu)   \right|^2 \right) } \right] \nonumber 
\end{align}

To simplify the notation, we do not express the dependence with the frequency $\nu$ and we get:
\begin{align}
    g_{ij} &=\sum_{\nu}\dfrac{1}{\gamma_{0}^2} E_{\boldsymbol{\xi}}\left[  { \left( -2  Re \left\lbrace x^*{\partial_i} s_{\boldsymbol{\xi}} \right\rbrace \right)  }
{ \left( -2  Re \left\lbrace x^* {\partial_j} s_{\boldsymbol{\xi}}  \right\rbrace  \right) } \right] \nonumber \\
&+ \sum_{\nu}\dfrac{1}{\gamma_{0}^2} E_{\boldsymbol{\xi}}\left[  { \left( -2  Re \left\lbrace x^*{\partial_i} s_{\boldsymbol{\xi}} \right\rbrace \right)  }
{ \left(  {\partial_j}  \left| s_{\boldsymbol{\xi}}   \right|^2 \right) } \right] \nonumber \\
&+\sum_{\nu}\dfrac{1}{\gamma_{0}^2} E_{\boldsymbol{\xi}}\left[  { \left(  {\partial_i}\left| s_{\boldsymbol{\xi}}    \right|^2 \right)  }
{ \left( -2  Re \left\lbrace x^* {\partial_j} s_{\boldsymbol{\xi}}  \right\rbrace  \right) } \right] \nonumber \\
&+\sum_{\nu}\dfrac{1}{\gamma_{0}^2} E_{\boldsymbol{\xi}}\left[  { \left(  {\partial_i}\left| s_{\boldsymbol{\xi}}   \right|^2 \right)  }
{ \left(  {\partial_j}  \left| s_{\boldsymbol{\xi}}    \right|^2 \right) } \right] \nonumber 
\end{align}
With expectation $E_{\boldsymbol{\xi}}\left[ ...\right]$, we obtain:
\begin{align}
    g_{ij} &=\sum_{\nu}\dfrac{2}{\gamma_{0}}  {   Re \left\lbrace {\partial_i} s_{\boldsymbol{\xi}}^* ~{\partial_j} s_{\boldsymbol{\xi}}  \right\rbrace   }  \nonumber
\end{align}
\\
From equation $\ref{eq10} $, this can be rewritten as follows.
\begin{equation}
g_{ij}=\sum_{\nu} \dfrac{2}{\gamma_{0}} Re \left\lbrace \partial_{i}
\left[  \rho_{\boldsymbol{\phi}}  \cdot \exp \left( -\imath \psi_{\boldsymbol{\varphi}} \right) \right] 
\partial_{j} \left[  \rho_{\boldsymbol{\phi}}  \cdot \exp \left(\imath \psi_{\boldsymbol{\varphi}}   \right) \right]
\right\rbrace \nonumber \\
\label{eq13A}
\end{equation}
\\
And we get:
\begin{equation}
g_{uv}=g_{vu}=\sum_{\nu} \dfrac{2}{\gamma_{0}}  \partial_{u}
\rho_{\boldsymbol{\phi}}  
\partial_{v}  \rho_{\boldsymbol{\phi}}  
\label{eq14A}
\end{equation}
\\
\begin{equation}
g_{qr}= g_{rq}=\sum_{\nu} \dfrac{2}{\gamma_{0}} \rho_{\boldsymbol{\phi}}^{2} \cdot   \partial_{q}
\psi_{\boldsymbol{\varphi}} 
\cdot \partial_{r}  \psi_{\boldsymbol{\varphi}}  
\label{eq15A}
\end{equation}

\begin{equation}
g_{uq}= g_{qu}=0  \nonumber \\
\label{eq16A}
\end{equation}
\\
This means that the Fisher matrix is a block diagonal matrix:
\begin{equation}
\left[ g_{ij}\right]=
\begin{pmatrix}
\left[ g_{uv}\right]  & \left[ 0\right]  \\
\left[ 0\right] & \left[ g_{qr}\right] \\ 
\end{pmatrix} \label{eq17A} \nonumber \\
\end{equation}
\\
In addition, we remark that:
\begin{equation}
\partial_{q} g_{uv}=\partial_{r} g_{uv}=0   \\
\label{eq18A}
\end{equation}
And:
\begin{equation}
\partial_{u} g_{qr}=\partial_{u} g_{rq}=\sum_{\nu} \dfrac{4}{\gamma_{0}} \rho_{\boldsymbol{\phi}} \cdot \partial_{u}
\rho_{\boldsymbol{\phi}} \cdot   \partial_{q}
\psi_{\boldsymbol{\varphi}} 
\cdot \partial_{r}  \psi_{\boldsymbol{\varphi}}  \\
\label{eq19A}
\end{equation}
(with the same expression replacing $u$ by $v$.) \\
And also:
\begin{equation}
\partial_{u'} g_{uv}=\partial_{u'}g_{vu}=\sum_{\nu} \dfrac{2}{\gamma_{0}}  \partial_{u'u}^2
\rho_{\boldsymbol{\phi}} \partial_{v}  \rho_{\boldsymbol{\phi}} + \sum_{\nu} \dfrac{2}{\gamma_{0}}  \partial_{u} \rho_{\boldsymbol{\phi}} \partial_{u'v}^2  \rho_{\boldsymbol{\phi}}
\label{eq14AAA}
\end{equation}

\section{Equations giving the Christoffel symbols}\label{secB}

Combining equation $\ref{eq02}$ and equations $\ref{eq14A}$ $\ref{eq15A}$ , we derive the equations of the Christoffel symbols. First, we obtain two equations corresponding to the Christoffel symbols related to the two diagonal blocks of the Fisher matrix:
\begin{equation}
\forall \ u,u',v' =1,...P \ \ \    \Gamma_{u'v',u} := g_{uv} \Gamma_{u'v'}^{v} = \frac{1}{2} \left( \partial_{u'} g_{v'u}  +
\partial_{v'} g_{uu'} - \partial_{u} g_{u'v'} \right)   \label{eq20A} \nonumber \\
\end{equation} 

\begin{equation}
\forall \ q,q',r' =P+1,...N \ \ \     \Gamma_{q'r',q}:= g_{qr} \Gamma_{q'r'}^{r} = \frac{1}{2} \left( \partial_{q'} g_{r'q} +
\partial_{r'} g_{qq'} - \partial_{q} g_{q'r'} \right)   \label{eq21A} \nonumber \\
\end{equation}
\\
In addition, we must also consider the case where non zero Christoffel symbols are derived because $g_{qr}$ (which is related to the $\varphi$ parameters) depends on $\phi$. This leads to a derivation $\partial_{u} g_{qr}$ that is not zero:

\begin{equation}
\forall \ u=1,...P \ \ \ \forall \ q',r' =P+1,...N \ \ \    \Gamma_{q'r',u}:= g_{uv} \Gamma_{q'r'}^{v} = \frac{1}{2} \left(  - \partial_{u} g_{q'r'} \right)   \label{eq22A} \nonumber \\
\end{equation} 
\\
and also:

\begin{equation}
\forall \ u=1,...P \ \ \ \forall \ q,q' =P+1,...N \ \ \     \Gamma_{uq',q}:=g_{qr} \Gamma_{uq'}^{r} = \frac{1}{2} \left( \partial_{u} g_{q'q} \right)    \label{eq23A} 
\end{equation} 
\\
and:

\begin{equation}
\forall \ u=1,...P \ \ \ \forall \ q,q' =P+1,...N \ \ \     \Gamma_{q'u,q}:=g_{qr} \Gamma_{q'u}^{r} = \frac{1}{2} \left( \partial_{u} g_{qq'} \right)   \label{eq24A} 
\end{equation} 
We note that equation $\ref{eq24A} $ can be deduced from equation $\ref{eq23A} $, if we consider symmetry:
\begin{equation}
\forall \ i,j,k =1,...N \ \ \      \Gamma_{ij}^{k} =  \Gamma_{ji}^{k}  \label{eq25A} \nonumber \\
\end{equation}  

The total maximal number of non zero Christoffel symbols in the equations is equal to $P^3+(N-P)^3+3P \cdot (N-P)^2$. This number is to be compared with $N^3$, the total number of Christoffel symbols. As an example, if we use one parameter for the magnitude and two parameters for the phase, we have a total of 27 Christoffel symbols, with a maximum of 21 non zero symbols.
\\

From the expression of the block diagonal Fisher matrix, these equations can be written as two matrix linear equations:
\begin{equation}
\left[ g_{uv}\right] \left[ \Gamma_{u'v'}^v \mid \Gamma_{q'r'}^v \right] =\left[ \Gamma_{u'v',u} \mid \Gamma_{q'r',u}\right]
\label{eq26B} \nonumber \\
\end{equation}
\begin{equation}
\left[ g_{qr}\right] \left[ \Gamma_{q'r'}^r \mid \Gamma_{q'u}^r \mid \Gamma_{uq'}^r\right] 
=\left[ \Gamma_{q'r',q} \mid \Gamma_{q'u,q} \mid \Gamma_{uq',q} \right]
\label{eq27B} \nonumber \\
\end{equation}
\\
In these equations, the terms on the right are given by the following expressions (we indicate the dependence on $\nu$):\\

$\forall \ u,u',v' =1,...P \ \ \forall \ q,q',r' =P+1,...N$
\begin{equation}
\Gamma_{u'v',u}
=  \sum_{\nu \in \mathcal{B}} \dfrac{2}{\gamma_{0}(\nu)}  \partial_{u}
\rho_{\boldsymbol{\phi}}(\nu) \cdot  \partial_{u'v'}^2  \rho_{\boldsymbol{\phi}}(\nu) 
\label{eq28B} 
\end{equation}
\begin{equation}
\Gamma_{q'r',u} = -\sum_{\nu \in \mathcal{B}} \dfrac{2}{\gamma_{0}(\nu)} \rho_{\boldsymbol{\phi}}(\nu) \cdot \partial_{u}
\rho_{\boldsymbol{\phi}}(\nu) \cdot   \partial_{q'}
\psi_{\boldsymbol{\varphi}}(\nu) 
\cdot \partial_{r'}  \psi_{\boldsymbol{\varphi}}(\nu)  \label{eq29B} 
\end{equation} 
\begin{equation}
\Gamma_{q'r',q} = \sum_{\nu \in \mathcal{B}} \dfrac{2}{\gamma_{0}(\nu)} \left( \rho_{\boldsymbol{\phi}}(\nu) \right)^{2} \cdot   \partial_{q} \psi_{\boldsymbol{\varphi}}(\nu) 
\cdot \partial_{q'r'}^2  \psi_{\boldsymbol{\varphi}}(\nu)   \label{eq30B} 
\end{equation}
\\
\begin{equation}
\Gamma_{q'u,q}  =\Gamma_{uq',q}  = \sum_{\nu \in \mathcal{B}} \dfrac{2}{\gamma_{0}(\nu)} \rho_{\boldsymbol{\phi}}(\nu) \cdot \partial_{u}
\rho_{\boldsymbol{\phi}}(\nu) \cdot   \partial_{q}
\psi_{\boldsymbol{\varphi}}(\nu) 
\cdot \partial_{q'}  \psi_{\boldsymbol{\varphi}}(\nu)  \label{eq31B} 
\end{equation} 
\\

\begin{proof} [Proof of equation \ref{eq28B}.] - \\
We have:
\begin{equation}
\forall \ u,u',v' =1,...P \ \ \    \Gamma_{u'v',u} = \frac{1}{2} \left( \partial_{u'} g_{v'u}  +
\partial_{v'} g_{uu'} - \partial_{u} g_{u'v'} \right)    \nonumber \\
\end{equation}
And taking into account equation \ref{eq14AAA}:
\begin{align}
    \Gamma_{u'v',u} &=   \sum_{\nu} \dfrac{1}{\gamma_{0}}  \partial_{u'u}^2
\rho_{\boldsymbol{\phi}} \partial_{v'}  \rho_{\boldsymbol{\phi}} + \sum_{\nu} \dfrac{1}{\gamma_{0}}  \partial_{u} \rho_{\boldsymbol{\phi}} \partial_{u'v'}^2  \rho_{\boldsymbol{\phi}} \nonumber \\ 
&+\sum_{\nu} \dfrac{1}{\gamma_{0}}  \partial_{v'u}^2
\rho_{\boldsymbol{\phi}} \partial_{u'}  \rho_{\boldsymbol{\phi}} + \sum_{\nu} \dfrac{1}{\gamma_{0}}  \partial_{u} \rho_{\boldsymbol{\phi}} \partial_{v'u'}^2  \rho_{\boldsymbol{\phi}} \nonumber \\
 &- \sum_{\nu} \dfrac{1}{\gamma_{0}}  \partial_{u'u}^2
\rho_{\boldsymbol{\phi}} \partial_{v'}  \rho_{\boldsymbol{\phi}} - \sum_{\nu} \dfrac{1}{\gamma_{0}}  \partial_{u'} \rho_{\boldsymbol{\phi}} \partial_{uv'}^2  \rho_{\boldsymbol{\phi}}    \nonumber \\
&=\dfrac{2}{\gamma_{0}}  \partial_{u} \rho_{\boldsymbol{\phi}} \partial_{v'u'}^2  \rho_{\boldsymbol{\phi}} \nonumber 
\end{align}
\end{proof} 

The proofs of equations \ref{eq29B}, \ref{eq30B}, and \ref{eq31B} use the same technique and are left to the reader. \\

From the previous equations, we can deduce the following property:
\begin{equation}
\Gamma_{q'u,q}  =\Gamma_{uq',q} =-\Gamma_{q'q,u}   \nonumber \\
\end{equation}

\section{Geodesic equations expressed on the manifolds $L2(\mathcal{B},\xi)$}\label{secC0}

Taking into account the notational convention from section \ref{sec2}, the equations of geodesics \ref{eq03}  can be written as follows.

\begin{equation}
\forall \ u=1,...P \ \ \    \left[ g_{uv}\right] \frac{d^{2} \phi^{v}}{d\varsigma^{2}} +\Gamma_{u'v',u} \frac{d \phi^{u'}}{d\varsigma} 
\frac{d \phi^{v'}}{d\varsigma} + \Gamma_{q'r',u} \frac{d \varphi^{q'}}{d\varsigma} 
\frac{d \varphi^{r'}}{d\varsigma}= 0 \label{eq35C} 
\end{equation}

\begin{equation}
\forall \ q =P+1,...N \ \ \    \left[ g_{qr}\right] \frac{d^{2} \varphi^{r}}{d\varsigma^{2}} +\Gamma_{q'r',q} \frac{d \varphi^{q'}}{d\varsigma} 
\frac{d \varphi^{r'}}{d\varsigma} + 2 \cdot \Gamma_{q'u',q} \frac{d \varphi^{q'}}{d\varsigma} 
\frac{d \phi^{u'}}{d\varsigma} = 0 \label{eq36C} 
\end{equation}

Taking into account the definition of the derivation, we can remark that, for all $\nu \in \mathcal{B}$:

\begin{equation}
\partial_{u} \rho_{\boldsymbol{\phi}}(\nu) \cdot \frac{d \phi^{u}}{d\varsigma}= \frac{\partial \rho_{\boldsymbol{\phi}}}{\partial \phi^{u}}(\nu)
\frac{d \phi^{u}}{d\varsigma}= \frac{d \rho_{\boldsymbol{\phi}}}{d\varsigma}(\nu) \label{eq37C} \nonumber \\
\end{equation}

\begin{equation}
\partial_{r} \psi_{\boldsymbol{\varphi}}(\nu) \cdot \frac{d \varphi^{r}}{d\varsigma}= \frac{\partial \psi_{\boldsymbol{\varphi}}}{\partial \varphi^{r}}(\nu)
\frac{d \varphi^{r}}{d\varsigma}= \frac{d \psi_{\boldsymbol{\varphi}}}{d\varsigma} (\nu) \label{eq38C} \nonumber \\
\end{equation}
And:
\begin{equation}
\frac{d^{2} \rho_{\boldsymbol{\phi}}}{d\varsigma^{2}}(\nu)=\frac{d \left( \frac{d \rho_{\boldsymbol{\phi}}}{d\varsigma}(\nu)\right) }{d\varsigma}
=\partial_{v'}\left( \frac{d \rho_{\boldsymbol{\phi}}}{d\varsigma}(\nu)\right) \cdot \frac{d \phi^{v'}}{d\varsigma}
= \partial_{v'}\left( \partial_{u'}\rho_{\boldsymbol{\phi}}(\nu) \cdot \frac{d \phi^{u'}}{d\varsigma}\right) \cdot \frac{d \phi^{v'}}{d\varsigma}
\label{eq39C} \nonumber \\
\end{equation}
Wich gives:
\begin{equation}
\frac{d^{2} \rho_{\boldsymbol{\phi}}}{d\varsigma^{2}}(\nu)
=\partial^2_{v'u'}\rho_{\boldsymbol{\phi}} (\nu)\cdot \frac{d \phi^{v'}}{d\varsigma} \frac{d \phi^{u'}}{d\varsigma}+\partial_{u'}\rho_{\boldsymbol{\phi}} (\nu)\cdot 
\partial_{v'}\left( \frac{d \phi^{u'}}{d\varsigma}\right) \cdot \frac{d \phi^{v'}}{d\varsigma}
=\partial^2_{v'u'}\rho_{\boldsymbol{\phi}} (\nu)\cdot \frac{d \phi^{v'}}{d\varsigma} \frac{d \phi^{u'}}{d\varsigma}+\partial_{u'}\rho_{\boldsymbol{\phi}} (\nu)\cdot 
\frac{d^2 \phi^{u'} }{d\varsigma^2}
\label{eq40C} \nonumber \\
\end{equation}
And, in the same way, we can obtain the following equality:
\begin{equation}
\frac{d^{2} \psi_{\varphi}}{d\varsigma^{2}}(\nu)
=\partial^2_{q'r'}\psi_{\varphi}(\nu) \cdot \frac{d \varphi^{q'}}{d\varsigma} \frac{d \varphi^{r'}}{d\varsigma}+\partial_{q'}\psi_{\varphi} (\nu)\cdot \frac{d^2 \varphi^{q'} }{d\varsigma^2}
\label{eq41C} \nonumber 
\end{equation} \\

When replacing the expressions of $g_{ij}$ and $\Gamma_{ij,k}$ in equations \ref{eq35C} and \ref{eq36C} by their expressions from equations \ref{eq14A} \ref{eq15A} and \ref{eq28B} \ref{eq29B} \ref{eq30B} \ref{eq31B}   and using the previous equalities, we obtain:
\begin{equation}
\forall \ u=1,...P \ \ \  \sum_{\nu \in \mathcal{B}} \dfrac{2}{\gamma_{0}(\nu)} \partial_{u}\rho_{\boldsymbol{\phi}}(\nu)  
\left( \frac{d^{2} \rho_{\boldsymbol{\phi}}}{d\varsigma^{2}}(\nu) -\rho_{\boldsymbol{\phi}}(\nu) \left(\frac{d \psi_{\boldsymbol{\varphi}}}{d\varsigma}(\nu) \right)^2  \right) 
= 0 \label{eq42C} \nonumber \\
\end{equation}

\begin{equation}
\forall \ q =P+1,...N \ \ \    \sum_{\nu \in \mathcal{B}} \dfrac{2}{\gamma_{0}(\nu)} \partial_{q}\psi_{\boldsymbol{\varphi}} (\nu) 
\left( \left(\rho_{\boldsymbol{\phi}}(\nu) \right)^2 \frac{d^{2} \psi_{\boldsymbol{\varphi}}}{d\varsigma^{2}}(\nu) 
+2 \rho_{\boldsymbol{\phi}}(\nu) \frac{d \rho_{\boldsymbol{\phi}}}{d\varsigma}(\nu)  \frac{d \psi_{\boldsymbol{\varphi}}}{d\varsigma} (\nu)  \right) 
= 0 \label{eq43C} \nonumber \\
\end{equation}

These two equations can be rewritten with the gradients $\boldsymbol{\bigtriangledown}_{\boldsymbol{\phi}} \rho_{\boldsymbol{\phi}}$ and $\boldsymbol{\bigtriangledown}_{\boldsymbol{\varphi}} \psi_{\boldsymbol{\varphi}}$:
\begin{equation}
\sum_{\nu \in \mathcal{B}} \dfrac{2}{\gamma_{0}(\nu)}  
\left( \frac{d^{2} \rho_{\boldsymbol{\phi}}}{d\varsigma^{2}}(\nu) -\rho_{\boldsymbol{\phi}}(\nu) \left(\frac{d \psi_{\boldsymbol{\varphi}}}{d\varsigma}(\nu) \right)^2  \right) 
\boldsymbol{\bigtriangledown}_{\boldsymbol{\phi}} \rho_{\boldsymbol{\phi}}(\nu) = \boldsymbol{\left[ 0\right]_{P}} 
\label{eq44C} \nonumber \\
\end{equation}

\begin{equation}
\sum_{\nu \in \mathcal{B}} \dfrac{2}{\gamma_{0}(\nu)}  
\left( \rho^2_{\boldsymbol{\phi}}(\nu) \frac{d^{2} \psi_{\boldsymbol{\varphi}}}{d\varsigma^{2}}(\nu) 
+2 \rho_{\boldsymbol{\phi}}(\nu) \frac{d \rho_{\boldsymbol{\phi}}}{d\varsigma}(\nu)  \frac{d \psi_{\boldsymbol{\varphi}}}{d\varsigma} (\nu)  \right) 
\boldsymbol{\bigtriangledown}_{\boldsymbol{\varphi}} \psi_{\boldsymbol{\varphi}}(\nu) = \boldsymbol{\left[ 0\right]_{N-P}}
\label{eq45C} \nonumber \\
\end{equation} \\

\section{Solutions of the geodesic equations on the submanifold $L2(\mathcal{B},\alpha)$}\label{secD}

We first have to consider the following equation.

\begin{equation}
\frac{d^{2} \alpha} {d\varsigma^{2}} 
= \frac{K}{\alpha^3}  
\label{eq60D1}
\end{equation}

The general solutions are as follows (with $\alpha \ge 0$).
\begin{equation}
\tilde{\alpha}(\varsigma) 
= \frac{\sqrt{k^2_1 \varsigma^{2}+2 k_2 k^2_1 \varsigma+k^2_2 k^2_1 +K}}{\sqrt{k_1}}
= {\sqrt{k_1 (\varsigma+ k_2)^2  +\frac{K}{k_1}}} \label{eq62DDD}
\end{equation}
which gives:
\begin{equation}
\frac{d\tilde{\alpha}}{d\varsigma}(\varsigma)
= \frac{k_1 \sqrt{k_1} (k_2+\varsigma)}{ \sqrt{k^2_1 \varsigma^{2}+2 k_2 k^2_1 \varsigma+k^2_2 k^2_1 +K} }  \label{eq63D}
\end{equation}

To determine the values of $k_1$ and $k_2$, we use the following equations:
\begin{equation}
\tilde{\alpha}(0)=\alpha_1
\end{equation}
\begin{equation}
\tilde{\alpha}(1)=\alpha_2
\end{equation}
\\
These give the following.
\begin{equation}
\alpha^2_2 k_1=k^2_1 +2 k_2 k^2_1 +k^2_2 k^2_1 +K
\end{equation}
\begin{equation}
\alpha^2_1 k_1=k^2_2 k^2_1 +K
\end{equation}
\\

We then look at the second equation:
\begin{equation}
\forall \nu \in \mathcal{B} \ \ \ \ \ \frac{d \psi_{\boldsymbol{\varphi}}}{d\varsigma} 
= \frac{1}{\alpha^2} \cdot  {c(\nu)}   \label{eq25E}
\end{equation} 

Based on equations \ref{eq62DDD} we have:
\begin{equation}
\forall \nu \in \mathcal{B} \ \ \ \ \ \frac{d \psi_{\boldsymbol{\varphi}}}{d\varsigma} 
= \frac{c(\nu) k_1 }{{{k_1^2 (\varsigma+ k_2)^2  +{K}}}}      \label{eq25F}
\end{equation}

Integrating and taking into account the boundary conditions $\psi_{\boldsymbol{\varphi}} \left( 0 \right) =\psi_{\boldsymbol{\varphi_{1}}}$, $\psi_{\boldsymbol{\varphi}} \left( 1 \right) =\psi_{\boldsymbol{\varphi_{2}}}$, with $\Delta \psi_{\boldsymbol{\varphi}}(\nu)=\psi_{\boldsymbol{\varphi}_2}(\nu)-\psi_{\boldsymbol{\varphi}_1}(\nu)$, we get:
\begin{equation}
\forall \nu \in \mathcal{B} \ \ \ \ \ c(\nu)= \sqrt{K}\dfrac {\Delta \psi_{\boldsymbol{\varphi}}(\nu)}
{\arctan \left( \frac{ {k_1}}{\sqrt{K}}  \left( k_2 +1 \right) \right) 
	-\arctan \left( \frac{ {k_1}}{\sqrt{K}}  \left( k_2  \right) \right) }
\end{equation}

We recall the definition of $K$: $K= \frac{1}{\omega_{0}} 
\sum_{\nu \in \mathcal{B}} \dfrac{2}{\gamma_{0}(\nu)} \left(\rho_{0}(\nu)\right)^2{\left( c(\nu)\right)^2 }$.\\

We may consider $\delta$ the weighted mean square difference of the phases (between the points $\psi_{\boldsymbol{\varphi}_1}$ and $\psi_{\boldsymbol{\varphi}_2}$): $\delta=\sqrt{\frac{1}{\omega_{0}} 
	\sum_{\nu \in \mathcal{B}} \dfrac{2}{\gamma_{0}(\nu)} \left(\rho_{0}(\nu) \right)^2{\left( \Delta \psi_{\boldsymbol{\varphi}}(\nu)\right)^2 }}$. \\
And we obtain the equation:
\begin{equation}
\left( \arctan \left( \frac{ {k_1}}{\sqrt{K}}  \left( k_2 +1 \right) \right) 
-\arctan \left( \frac{ {k_1}}{\sqrt{K}}  \left( k_2  \right) \right) \right)^2 
= \left( \delta \right)^2
\end{equation}

In order to compute the parameters $k_1$, $k_2$, $K$, we have to solve the equations:
\begin{align}
& \alpha^2_2 -\alpha^2_1  = k_1 +2 k_2 k_1 \\
& K = \alpha^2_1 k_1-k^2_2 k^2_1   \label{eq76DD} \\
& \arctan \left( \frac{ {k_1}}{\sqrt{K}}  \left( k_2 +1 \right) \right) 
-\arctan \left( \frac{ {k_1}}{\sqrt{K}}  \left( k_2  \right) \right)  
= \delta \label{eq77DD} \
\end{align}

Using the formula $\tan \left(a-b \right)=\left( \tan a - \tan b\right) / (1+ \tan a \tan b) $ on the last equation \ref{eq77DD}, and then replacing K from equation \ref{eq76DD}, we get:
\begin{align*}
\tan \delta & = \dfrac{ \frac{ {k_1}}{\sqrt{K}}  \left( k_2 +1 \right) -  \frac{ {k_1}}{\sqrt{K}}  \left( k_2  \right)}
{1+\frac{ {k_1}}{\sqrt{K}}  \left( k_2 +1 \right) \cdot \frac{ {k_1}}{\sqrt{K}}  \left( k_2  \right)} \\
& = \dfrac{ \frac{ {k_1}}{\sqrt{K}}  }
{1+\frac{ (k_1)^2}{K}  \left( k_2 +1 \right) \cdot   \left( k_2  \right)} \\
& = \dfrac{  k_1 \sqrt{K} }
{ K+(k_1)^2   k_2^2 +(k_1)^2 k_2 } \\
& = \dfrac{  k_1 \sqrt{K} }
{ \alpha^2_1 k_1-k^2_2 k^2_1 +(k_1)^2   k_2^2 +(k_1)^2 k_2 } \\
& = \dfrac{  \sqrt{K} }
{ \alpha^2_1  +(k_1) k_2 } \\
& = \dfrac{   \sqrt{\alpha^2_1 k_1-k^2_2 k^2_1} }
{ \alpha^2_1 +(k_1) k_2 }
\end{align*}

Replacing $k_2$ with its value depending on $k_1$ and $\alpha_{1}$, $\alpha_{2}$, we obtain the equations:
\begin{equation}
\left( \tan \delta \right)^2 \left( (\alpha_1)^2+(\alpha_2)^2 - k_1\right)^2
+  \left( (\alpha_2)^2-(\alpha_1)^2 - k_1\right)^2 - 4 {(\alpha_1)^2}{k_1}=0
\label{eq78D}
\end{equation}
\begin{equation}
k_2= \dfrac{(\alpha_2)^2 - (\alpha_1)^2}{2 k_1} - \frac{1}{2}
\label{eq79D}
\end{equation}
\begin{equation}
K= (\alpha_1)^2 k_1  - \frac{1}{4} \left((\alpha_2)^2-(\alpha_1)^2 - k_1 \right)^2 
\label{eq80D}
\end{equation}
\\
The solutions are :
\begin{equation}
k_1= (\alpha_2)^2 + (\alpha_1)^2 - 2 \alpha_1 \alpha_2 \cos \delta
\label{eq81D}
\end{equation}
\begin{equation}
k_2= \dfrac{- (\alpha_1)^2 + \alpha_1 \alpha_2 \cos \delta}{(\alpha_2)^2 + (\alpha_1)^2 - 2 \alpha_1 \alpha_2 \cos \delta} 
\label{eq82D}
\end{equation}
\begin{equation}
K= (\alpha_1)^2 (\alpha_2)^2 (\sin \delta)^2 
\label{eq83D}
\end{equation}

\end{appendices}

%%===========================================================================================%%
%% If you are submitting to one of the Nature Portfolio journals, using the eJP submission   %%
%% system, please include the references within the manuscript file itself. You may do this  %%
%% by copying the reference list from your .bbl file, paste it into the main manuscript .tex %%
%% file, and delete the associated \verb+\bibliography+ commands.                            %%
%%===========================================================================================%%

\bibliography{sn-bibliography}% common bib file
%% if required, the content of .bbl file can be included here once bbl is generated
%%\input sn-article.bbl

\end{document}